\documentclass[aps,onecolumn,11pt,superscriptaddress,nofootinbib]{revtex4}

\usepackage{graphicx}
\usepackage{amsmath,amssymb,bm,physics,ulem}
\usepackage{autobreak,mathtools}
\usepackage{comment}
\usepackage[dvipsnames, usenames]{xcolor}
\usepackage[scr=boondoxo,scrscaled=1.05]{mathalfa}

\definecolor{tealgreen}{rgb}{0.0, 0.5, 1.0}
\definecolor{darkblue}{rgb}{0., 0.4, 0.8}
\definecolor{treegreen}{rgb}{0., 0.7, 0.3}
\definecolor{cadmiumred}{rgb}{1., 0., 0.22}
\definecolor{orchid}{rgb}{0.7., 0., 0.5}
 
\usepackage[bookmarks,linktocpage, colorlinks=true, plainpages = false, citecolor = treegreen,  linkcolor=darkblue, urlcolor = darkblue, filecolor = blue]{hyperref}

\definecolor{changes}{RGB}{169.2, 155.0, 0.5}

\def\be#1\ee{\begin{align}#1\end{align}}

\def\ba{\begin{eqnarray}}
	\def\ea{\end{eqnarray}}
\def\nn{\nonumber}

\begin{document}

\title{Effective geometrodynamics for renormalization-group improved black-hole spacetimes in spherical symmetry}

\author{Johanna Borissova}
\email{j.borissova@imperial.ac.uk}
\affiliation{Abdus Salam Centre for Theoretical Physics, Imperial College London, London SW7 2AZ, UK}
\author{Ra\'ul Carballo-Rubio}
\email{raul.carballorubio@iaa.csic.es}
\affiliation{Instituto de Astrof\'isica de Andaluc\'ia (IAA-CSIC),
Glorieta de la Astronom\'ia, 18008 Granada, Spain}

\begin{abstract}
	\bigskip 
{\sc Abstract:}
We consider the spherically reduced Einstein--Hilbert action, Einstein field equations and Schwarzschild spacetime modified by a renormalization-group (RG) scale-dependent gravitational Newton coupling, and present a systematic and operational approach to such an RG-improvement. The master field equations for spherically symmetric gravitational fields, recently constructed from two-dimensional Horndeski theory, allow us to retain partial contributions from higher-curvature truncations of the effective action, while preserving the second-order nature of the resulting field equations. Static RG-improved black-hole  spacetimes with an effective gravitational coupling depending on the areal radius and the Misner-Sharp mass are derived as vacuum solutions to these master field equations, and are thereby identified as solutions to generally covariant two-dimensional Horndeski theories. We discuss explicitly the embedding of previous key works on RG-improvement into the newly developed formalism to illustrate its broad range of applicability. This formalism moreover allows us to establish explicitly the discrepancies in the outcomes of RG-improvement when implemented at the level of the action, in the field equations, or in the Schwarzschild solution.
\end{abstract}

\maketitle

\tableofcontents

\section{Introduction}

General relativity provides a remarkably accurate description of gravitational phenomena in the large-scale regime of strong gravitational fields~\cite{Will:2014kxa,LIGOScientific:2020tif,LIGOScientific:2021sio}. At small scales, however, this theory is yet to be replaced by an ultraviolet completion of quantum gravity and matter. Viable candidates are expected in particular to feature mechanisms counteracting the formation of classical singularities~\cite{Carballo-Rubio:2025fnc,Buoninfante:2024yth,Buoninfante:2024oxl,Bambi:2025wjx}. Singularity avoidance may result, e.g., from higher-derivative and non-local terms entering the gravitational effective action~\cite{Buoninfante:2022ild,Platania:2023srt,Eichhorn:2022bgu}.

Adopting this viewpoint, several key challenges arise. On the one hand, deriving the effective action from a given quantum-gravity theory is far from trivial. To circumvent this challenge, criteria for filtering out effective actions from a landscape of effective field theories have been formulated, such as positivity bounds on generic effective field theories~\cite{Adams:2006sv,Bellazzini:2020cot,Tolley:2020gtv}, and the swampland criteria in string theory~\cite{Vafa:2005ui,Ooguri:2006in,Palti:2019pca,vanBeest:2021lhn}. Applications of these criteria have been recently extended into the context of asymptotic safety~\cite{Basile:2021krr,Knorr:2024yiu,Eichhorn:2024wba}. Whether or not the underlying  criteria are physically viable, and whether or not the resulting theory spaces have an overlap of non-zero measure, is subject to ongoing debate~\cite{Eichhorn:2024rkc,Basile:2025zjc,Borissova:2024hkc}. On the other hand, regardless of the specific form of the effective action, it is widely expected that corrections to general relativity manifest themselves through higher-derivative operators, resulting in higher-than-second-order non-linear field equations.  Such differential equations are difficult to solve in general, and moreover solutions are expected to exhibit classical instabilities in the dynamical evolution~\cite{Ostrogradsky:1850fid,deUrries:1998obu,Woodard:2015zca}.~\footnote{See, however,~\cite{Salvio:2019ewf,Deffayet:2021nnt,Deffayet:2023wdg,ErrastiDiez:2024hfq,Deffayet:2025lnj,Held:2025fii} for results indicating that there exist classical dynamical systems and field theories with unbounded Hamiltonians which can still exhibit long-lived stable dynamics.} Thus, even with an effective field theory beyond general relativity at hand, the task of deriving quantum-gravitational corrections to classical spacetimes remains a fundamental challenge.~\footnote{Since the pioneering classification of static spherically symmetric solutions in quadratic gravity~\cite{Stelle:1977ry,Lu:2015cqa,Lu:2015psa,Podolsky:2018pfe,Podolsky:2019gro}, there have been only few attempts of extending such analyses to higher-derivative theories. Examples are six-derivative gravity~\cite{Giacchini:2024exc,Giacchini:2025gzw} and quasi-local Einstein--Weyl gravity~\cite{Borissova:2025nvj}.}

Tackling this challenge in full generality is at present out of reach. In this context, highly symmetric frameworks oftentimes provide a playground in which explicit computations can be performed, thereby allowing for the identification of potential roadblocks of an approach to the full problem. Gaining qualitative insights by resorting to symmetry-reduced analyses on lower-dimensional configuration spaces is often referred to as ``minisuperspace", following the widely known application of such an approach in the context of quantum cosmology~\cite{DeWitt:1967yk,Hartle:1983ai,Vilenkin:1986cy}. In this spirit, our main goal here is to exploit the power of symmetry reduction, in order to establish a pathway connecting two of the aforementioned challenges: the inevitability of higher-derivative operators for an accurate description of expected quantum-gravitational phenomena, combined with the challenge of solving higher-order field equations.

Concretely, we will focus on quantum-gravitational effects on black-hole spacetimes expected as a result of an ultraviolet fixed point of the dimensionless gravitational Newton coupling according to asymptotic safety~\cite{Weinberg:1976xy,Eichhorn:2018yfc,Niedermaier:2006wt}. Such effects can be studied qualitatively by means of a procedure referred to as renormalization-group (RG) improvement~\cite{Eichhorn:2022bgu,Platania:2023srt}, whereby classical couplings are replaced by scale-dependent couplings depending on the RG momentum, which is subsequently promoted to a position-dependent quantity capturing the physical energy scales of a given system. This procedure provides a playground for exploring the mechanism of singularity resolution,  resulting intuitively from the weakening of the dimensionful Newton coupling at high energies, while avoiding the task of dealing with higher-derivative and non-local terms at higher orders in the truncation. Yet, this procedure has a fundamental drawback manifested in ambiguities of the implementation, such as in the freedom of performing the replacement of couplings at the level of the action, field equations, or solutions, or in the choice of scale identification (see, e.g.,~\cite{Borissova:2022mgd} for a discussion on these issues), as well as in the dependence on the coordinate system in which the replacement at the level of solutions is implemented~\cite{Held:2021vwd}, and finally in the difficulty of addressing how the outcomes of different implementations are connected. In particular, the straightforward step of RG-improving the singular Schwarzschild solution of general relativity, into a regular black-hole spacetime, cf.~originally~\cite{Bonanno:2000ep}, loses manifest connection to a variational principle. In other words, it is unclear whether the resulting spacetimes can arise as actual solutions to a generally covariant gravitational theory. The same applies to an RG-improvement at the level of the Einstein field equations. 

In this work, we will reconcile the aforementioned notions and implementations of RG-improvement within the subspace of spherically symmetric geometries. To that end, we will resort to a recently constructed ``effective geometrodynamic'' framework~\cite{Carballo-Rubio:2025ntd}, in which a set of master field equations generalizing the Einstein field equations on spherically symmetric backgrounds has been constructed from two-dimensional Horndeski theory~\cite{Horndeski:1974wa,Kobayashi:2011nu,Kobayashi:2019hrl}. We will demonstrate that these master field equations are general enough to describe spacetimes obtained from an RG-improvement of the static Schwarzschild spacetime, as effective vacuum solutions, and can be used to model the dynamical evolution into such geometries when coupled to an energy-momentum tensor for collapsing matter. Moreover, we will discuss conditions under which the RG-improvement procedure yields gravitational actions that can be written as two-dimensional Horndeski theories, and discuss how the required truncations can provide a manifest and systematic way to take into account partial contributions of  higher-curvature terms in the effective action, while preserving the second-order nature of the equations of motion.

The remainder of this article is structured as follows. Section~\ref{Sec:Overview} contains a detailed overview on the motivations and applications of RG-improvement to black-hole physics. Concretely, Subsection~\ref{SecSub:RunningG} explains the idea of incorporating an RG scale-dependent Newton coupling into the classical theory. The procedures of RG-improvement at the level of classical solutions, field equations and actions are reviewed  in Subsections~\ref{SecSub:RGSols},~\ref{SecSub:RGEqs}, and~\ref{SecSub:RGActions}, respectively. Section~\ref{Sec:EffectiveGeometrodynamics} is devoted to the embedding of these procedures, performed within the subspace of spherically symmetric geometries,  into the effective geometrodynamic framework based on two-dimensional Horndeski theory, in which the aforementioned master field equations for spherically symmetric gravitational fields are formulated. To that end, concretely, Subsection~\ref{SecSub:2dHorndeski} contains an overview of two-dimensional Horndeski theory and therefrom constructed generalized Einstein field equations, which serve as the master equations for subsequent analyses. In Subsections~\ref{SecSub:RGImprovementSolutions} and~\ref{SecSub:RGImprovementEquationsActions}, we apply this framework to static RG-improved Schwarzschild spacetimes, the RG-improved spherically symmetric Einstein field equations and the RG-improved spherically reduced Einstein--Hilbert action. In Section~\ref{Sec:GravCollapse}, we describe the dynamical  gravitational collapse of classical matter into RG-improved Schwarzschild spacetimes. Section~\ref{Sec:ScaleIdentification} contains a discussion of admissible scale identifications and required truncations resulting in a two-dimensional Horndeski theory at the level of the full non-reduced RG-improved Einstein--Hilbert action. We finish with a discussion in Section~\ref{Sec:Discussion}.

\section{Overview of RG-improvement accounting for quantum-gravitational effects}\label{Sec:Overview}

\subsection{Running gravitational coupling}\label{SecSub:RunningG}

\begin{figure}[t]
	\centering
	\includegraphics[width=0.9\textwidth]{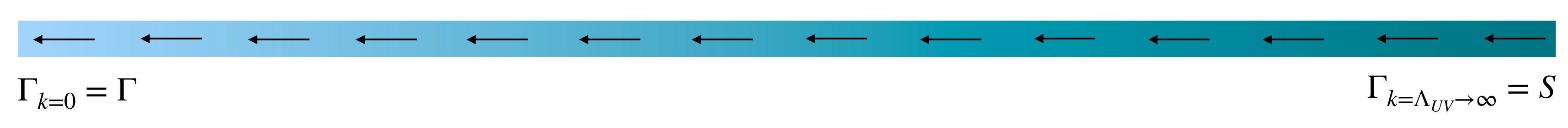}
	\caption{\label{Fig:Gammak} Effective average action $\Gamma_k$ interpolating between the bare action $S$ for $k \to \infty$ and the quantum effective action $\Gamma$ for $k \to 0$.}
\end{figure}

The dynamics in a  quantum field theory is governed by the quantum effective action $\Gamma$. The latter can for instance be computed using the functional renormalization group (RG) from an exact flow equation for an effective average action functional $\Gamma_k$ which interpolates between the bare action $S$ for $k\to \Lambda_{\text{UV}}$, where $\Lambda_{\text{UV}}$ denotes an ultraviolet (UV) cutoff scale, and the quantum effective action $\Gamma$ for $k \to 0$, where all quantum fluctuations have been integrated out~\cite{Wetterich:1992yh,Reuter:1992uk,Reuter:1993kw}, cf.~Figure~\ref{Fig:Gammak}. The RG scale parameter $k$ with canonical mass dimension of momentum, i.e., $[k] = +1$~\footnote{We adopt units such that $c = \hbar = 1$.}, plays the role of an effective infrared (IR) cutoff in the Wilsonian shell-by-shell integration of high-momentum modes. According to Weinberg's asymptotic safety~\cite{Weinberg:1976xy,Weinberg:1980gg}, a renormalizable quantum field theory is obtained if the RG flow in the theory space of couplings approaches a UV fixed point with a finite-dimensional UV critical surface. At such a fixed point the theory becomes quantum scale invariant, such that the limit $\Lambda_{\text{UV}} \to \infty$ can be taken. For reviews on asymptotically safe quantum gravity, see e.g.~\cite{Eichhorn:2022gku,Eichhorn:2018yfc,Percacci:2007sz,Niedermaier:2006wt}.
 
In practice, computing the quantum effective action by solving the functional RG equation is highly challenging and requires truncations of the theory space. The so-called RG-improvement procedure in asymptotically safe gravity, see e.g.~\cite{Eichhorn:2022bgu,Platania:2023srt} for reviews, provides a pathway to qualitatively analysing possible effects of quantum fluctuations on classical spacetimes by mimicking higher-order and non-local operators which have not been included in such a truncation. See e.g.~\cite{Bonanno:2000ep,Reuter:2003ca,Bonanno:2006eu,Falls:2010he,Torres:2014gta,Koch:2013owa,Kofinas:2015sna,Bonanno:2016dyv,Bonanno:2017zen,Bonanno:2017kta,Pawlowski:2018swz,Adeifeoba:2018ydh,Held:2019xde,Platania:2019kyx,Borissova:2022mgd,Borissova:2022jqj,Bonanno:2023rzk,Bonanno:2024paf,Bonanno:2024wvb} for applications to black-hole physics. The procedure consists in replacing classical couplings with their RG scale-dependent versions and subsequently relating the RG scale parameter to a characteristic physical energy by performing the scale identification $k \mapsto k(x)$, which makes the RG scale momentum a position-dependent quantity. RG-improvement can be implemented at different levels: in the action~\cite{Reuter:2003ca,Reuter:2004nv,Borissova:2022jqj,Bonanno:2024paf}, in the field equations~\cite{Bonanno:2002zb,Babic:2004ev,Kofinas:2015sna,Platania:2019kyx,Borissova:2022mgd}, or in the solutions~\cite{Bonanno:2000ep,Bonanno:2006eu,Falls:2010he,Koch:2013owa,Torres:2014gta,Bonanno:2016dyv,Bonanno:2017zen,Bonanno:2017kta,Pawlowski:2018swz,Adeifeoba:2018ydh,Platania:2019kyx,Held:2019xde,Borissova:2022mgd,Bonanno:2024wvb,Boos:2023xoq}  to a given classical theory.  The latter is the approach considered for gravity originally in~\cite{Bonanno:2000ep}.  
The viability of an RG-improvement procedure at the level of the action can be motivated by the decoupling mechanism in effective field theories~\cite{Reuter:2003ca,Borissova:2022mgd}. The spacetimes obtained as a result of RG-improvement can be viewed as effective spacetimes that may account for quantum-gravitational effects. These spacetimes may or may not be singularity-free, depending on the specifics of the implementation and on the couplings subjected to RG-improvement. For conditions on black-hole singularity resolution in asymptotically safe gravity, see e.g.~\cite{Adeifeoba:2018ydh}.

Before we proceed, let us elaborate on the notion of spacetime singularity used here. In the original context of the Penrose-Hawking singularity theorems~\cite{Penrose:1964wq,Hawking:1970zqf,Hawking:1973uf}, the notion of spacetime singularity is tied to the concept of  geodesic incompleteness of an inextendible manifold and as such is associated with a breakdown of predictivity in the dynamical evolution. Here, we will instead refer to a spacetime singularity as a point at which local curvature invariants are infinite, thereby indicating the divergence of local tidal forces. Both concepts of spacetime singularity are not equivalent as emphasised in~\cite{Wald:1984rg} (see also the related discussion in~\cite{Carballo-Rubio:2025fnc}). In particular, it is possible, e.g., that curvature invariants diverge simply on approach of infinity, or that curvature invariants are finite whereas components of the curvature tensor diverge, or that a spacetime has physically pathological properties even though the curvature tensor is well-behaved. Related to these aspects, in~\cite{Lukash:2011hd,Lukash:2013ts} the notion of an integrable singularity has been introduced by requiring that metric components are finite such that, e.g., even though a local curvature invariant diverges, its spacetime integral may remain finite. As pointed out in~\cite{Arrechea:2025fkk}, however, integrable singularities are focusing points which are reached at finite value of the affine parameter along a geodesic and are in addition generically unstable under perturbations. Moreover, physical quantities measured by a local observer may still accumulate in value in a way that prevents this observer from traversing the singularity. It is therefore questionable whether spacetimes with an integrable singularity are physically viable.\\

In the following, we will focus on an RG-improvement involving the classical Newton coupling $G_0$, which has canonical mass dimension  $[G_0] = -2$. In an RG-improvement, $G_0$ is replaced by the scale-dependent Newton coupling $G_k$, for which an interpolating function is given as follows~\cite{Bonanno:2000ep},
\be\label{eq:Gk}
G_0\quad \to\quad G_k \,;\,\,\,\quad  \quad \,\,\, G_k = \frac{G_0}{1+ G_0 \omega k^2}\,\,\,\quad  \text{with} \quad \,\,\, k=k(x)\,. 
\ee
Here $\omega = g_\star^{-1}$ denotes the inverse fixed-point value of the dimensionless Newton coupling $g_k = k^2 G_k$. The function~\eqref{eq:Gk} interpolates smoothly between the IR regime $k \to 0$, where $G_k$ reduces to the classical Newton constant $G_0$, and the UV regime $k \to \infty$, where $G_k$ scales as $k^{-2}$ according to its canonical mass dimension in the presence of a UV fixed point.

\subsection{RG-improvement of classical solutions}\label{SecSub:RGSols}

In this subsection, we review how effective spacetimes are obtained by an application of the RG-improvement procedure~\eqref{eq:Gk} at the level of classical solutions to general relativity.
 
For concreteness, we consider static spherically symmetric spacetimes with line element in spherical coordinates $(t,r,\theta,\phi)$ 
\be\label{eq:MetricSphericalSymmetry}
\dd{s^2} = -f(r)\dd{t}^2 + \frac{\dd{r}^2}{f(r)}+ r^2   \dd{\Omega}^2,
\ee
where $ \dd{\Omega}^2 = \dd{\theta}^2 + \sin^2\theta\dd{\phi}^2$ denotes the area element on the unit two-sphere. While the most general line element compatible with these symmetries is parametrised by two free functions such that $g_{tt}g_{rr}\neq 1$ in general, the line element above with a single function is enough to describe an RG-improvement of the Schwarzschild vacuum solution to general relativity.
In another classical theory in which $G_0$ appears as a coupling in static spherically symmetric solutions with two distinct metric functions, one would have to RG-improve $G_0 \to G_k$ separately in both metric functions, whereby the RG scale-dependence $G_k=G(k)$ would have to be derived from a computation of RG flows for this particular theory as a quantum field theory.

The Schwarzschild spacetime with lapse function given by
\be\label{eq:fSchwarzschild}
%f(r)=
f_0(r) = 1- \frac{2 M G_0}{r}
\ee
is the unique vacuum solution to the Einstein field equations under the assumption of spherical symmetry. The parameter $M$ with canonical mass dimension $[M] = +1$ is proportional to the Arnowitt--Deser--Misner (ADM) mass~\cite{Arnowitt:1959ah,Arnowitt:1962hi}, defined for asymptotically flat spacetimes. Considering charged or rotating black holes is out of the scope of this paper, although finding connections of the discussion below with related works~\cite{Ishibashi:2021kmf,Chen:2022xjk,Chen:2023wdg,Hao:2025utc} may be worth exploring.

In an RG-improvement of classical solutions, common choices of scale identification consist in relating the scale parameter $k$ to local-curvature or proper-distance scales, see, e.g.~\cite{Bonanno:2000ep,Bonanno:2006eu,Falls:2010he,Koch:2013owa,Pawlowski:2018swz,Adeifeoba:2018ydh,Held:2019xde}. For the Schwarzschild spacetime~\eqref{eq:fSchwarzschild}, the following type of scale identification has been considered originally in~\cite{Bonanno:2000ep},
\ba\label{eq:k1}
k^2\qty(r;M)   &=& \frac{G_0 M}{r^3} \,\,\, \quad \propto \quad \,\,\, \eval{\qty{\mathcal{K}^{\frac{1}{2}}\,, \frac{1}{\delta^2}}}_{\text{Schwarzschild}}\,,
\ea
where $\mathcal{K} = R_{\mu\nu\rho\sigma}R^{\mu\nu\rho\sigma}$ denotes the Kretschmann scalar and $\delta(r)$ the radial proper distance to the center. On dimensional grounds, one may more generally consider a scale identification as follows, see e.g.~\cite{Falls:2010he,Adeifeoba:2018ydh},
\ba\label{eq:k1pq}
k_q^2(r;M) &=& \frac{\qty(G_0 M)^{p}}{r^q}\,,
\ea
where $p= q-2$ and $q > 0$ is assumed.~\footnote{One may also consider more complicated scale identifications such as sums of terms~\eqref{eq:k1pq} with different values of $q>0$.} This assumption ensures that the limit $k_q^2 \to 0$, associated with IR physics, is mapped onto the asymptotic region $r \to \infty$ of large distances from the radial center, whereas the limit $k_q^2 \to \infty$, associated with UV physics, is mapped onto the spacetime region $r \to 0$ surrounding the black-hole core. In view of the interpolating function~\eqref{eq:Gk}, such a scale identification translates into the following asymptotics of the effective Newton coupling $G_{k_q}$ as a function of $r$,
\ba
\lim_{r \to \infty} G_{k_q} \,\,\, &  \to & \,\,\, G_0\,,\label{eq:GLimitIR}\\
\lim_{r\to 0} G_{k_q} \,\,\, & \propto  & \,\,\, r^{-q}\,.\label{eq:GLimitUV}
\ea
The resulting RG-improved spacetimes with lapse function
\be\label{eq:fk}
f_{k_q}(r) = 1 - \frac{2 G_{k_q}M}{r} = 1 - \frac{2 G_0 M r^{q-1}}{r^q + G_0 \omega \qty(G_0 M)^p}
\ee
are thereby guaranteed to be asymptotically flat, with mass parameter $M$, and to expand locally near $r=0$ as
\be
f_{k_q}(r) = 1 - \frac{2 M}{\omega \qty(G_0 M)^p }  \, r^{q-1} + \dots\,.
\ee
For $q \geq 3$ local curvature invariants of the RG-improved spacetime at $r=0$ are finite, as can be verified by an explicit calculation of the Kretschmann scalar. This is not the case, e.g., for $q=2$, corresponding to the naive scale identification $k(r) \propto r^{-1}$ appropriate in flat spacetime. For the choice  $q=3$ implied by~\eqref{eq:k1}, such that $p=1$ in~\eqref{eq:k1pq}, the spacetime obtained originally in~\cite{Bonanno:2000ep} as a result of RG-improvement of the Schwarzschild spacetime represents a regular Hayward spacetime~\cite{Hayward:2005gi} with an effective regularization length parameter given by $l \propto \sqrt[3]{G_0^2 \omega M }$.\\

RG-improvement at the level of the classical Schwarzschild solution results in an effective source on the right-hand side of the Einstein field equations. Concretely, any spacetimes~\eqref{eq:MetricSphericalSymmetry} with metric function of the form
\be\label{eq:RGimprovedSch}
f_k(r) = 1- \frac{2 M G_k(r)}{r}
\ee
can be described by the Einstein field equations written in the form
\ba
G_{\mu\nu} &=& -\Delta_k G_{\mu\nu}\,,
\ea
where $\Delta_k G_{\mu\nu}$ accounts for the modification of the classical spacetime as a result of RG-improving the Newton coupling, and is equivalent to an energy-momentum tensor  of the null-fluid type~\cite{Husain:1995bf,Wang:1998qx}, i.e.,
\be \label{eq:effsourcev}
\Delta_kG_{\mu\nu}=-8\pi G_k\qty[ \qty(\rho_k + p_k)\qty(l_\mu n_\nu + l_\nu n_\mu) + p_k g_{\mu\nu}]\,.
\ee
Here $l_\mu $ and $n_\mu$ are null vectors such that $l_\mu n^\mu = -1$. The effective energy density $\rho_k$ and pressure $p_k$ are determined by the first and second radial derivatives of the effective Newton coupling $G_k(r)$ as~\cite{Platania:2019kyx,Borissova:2022mgd}
\be \label{eq:RhoP}
\rho_k = \frac{M \partial_r G_k}{4 \pi r^2 G_k}\,\,\,\quad \text{and} \quad \,\,\,p_k =- \frac{M \partial_r^2 G_k}{8 \pi r G_k}\,.
\ee
This effective source contains up to second-order derivatives of $G_k$ with respect to the radial coordinate, as it is defined by calculating the Einstein tensor of the RG-improved spacetime. Hence, this definition of the effective source implicitly asserts that the modification of the Einstein field equations due to the RG-improvement of the Newton coupling in a static situation is of up to second order in derivatives, which will be an explicit assumption in the construction presented below. Specifically, we will show that any such effective source, accounting for a generic $r$-dependent Newton coupling $G_k(r)$ implemented in the static Schwarzschild metric, and including a possible dependence on the mass $M$, can be  treated  systematically in the framework of generalized second-order master field equations for spherically symmetric spacetimes, which are introduced in Section~\ref{SecSub:2dHorndeski}.

Before closing this subsection, let us mention a particular example of RG-improved Schwarzschild spacetimes to which the discussion presented in this paper can be applied. In general, the new energy scales generated as a result of RG-improving a given classical spacetime can be interpreted as an effective quantum-gravitational self-energy~\cite{Poisson:1988wc,Poisson:1990eh}, which induces a backreaction effect on the classical background. To account for such a backreaction, in~\cite{Platania:2019kyx} an iterated RG-improvement procedure has been proposed, starting from an RG-improved static Schwarzschild spacetime with an effective Newton coupling $G_{k_0}^{(1)}(r)$  obtained based on an initial scale identification $k_0 = k_0(r)$ using local curvature scales, which can be regarded as an initial perturbation of the classical system. Subsequently, repeated RG-improvements $G_{k_{n-1}}^{(n)} (r)\to G_{k_n}^{(n+1)}(r)$ are performed to obtain new spacetimes, whereby the scale identification $k_{n} = k_{n}(r)$ at each step of the iteration is constructed from the physical energy scales of the spacetime generated at the previous step. The discussion above shows that there will be in general, apart from local curvature scales, both an energy density $\rho_{n}(r)$ and a pressure $p_{n}(r)$ to be taken into account in the definition of the cutoff function. Under the assumption that the series of iterated RG-improved spacetimes converges, i.e., the limit $n\to \infty$ can be taken, the static spacetime at the fixed point of the iteration will be invariant under repeated RG-improvements. Its effective Newton coupling is obtained from the defining equation
\begin{equation}
	G^{(\infty)}_{k_\infty}(r)=\frac{G_0}{1+G_0\omega k_{\infty}^2\qty(G_{k_\infty}^{(\infty)},{G_{k_\infty}^{(\infty)}}',{G_{k_\infty}^{(\infty)}}''; r)}\,,   
\end{equation}
which contains up to second-order derivatives of $G_{k_\infty}^{(\infty)}$ itself, stemming from $\rho_{\infty}$ and $p_{\infty}$. This equation can be solved, at least in principle, once a specific form of the cutoff function $k_{n}^2$ used at each step of the iteration has been chosen. The result will be a function form $G_{k_\infty}^{(\infty)}(r)$ with a generally complicated dependence on the ADM mass parameter $M$. Later in Section~\ref{SecSub:RGImprovementSolutions}, we will describe how such spacetimes can be described within the effective geometrodynamic framework based on two-dimensional Horndeski theory which sets the ground for the extended Einstein equations constructed in~\cite{Carballo-Rubio:2025ntd}. The analogue treatment in appplications of RG-improvement to dynamical settings~\cite{Borissova:2022mgd} is beyond the scope of this paper, however, in Section~\ref{Sec:GravCollapse} we will discuss how the same framework can be applied to describe the classical gravitational collapse into RG-improved spacetimes of the general form~\eqref{eq:fk}.

\subsection{RG-improvement of classical equations}\label{SecSub:RGEqs}

In this subsection, we illustrate  the application of RG-improvement~\eqref{eq:Gk} at the level of the Einstein field equations in the presence of an external matter source $T_{\mu\nu}$,
\be\label{eq:EinsteinFieldEq}
\frac{G_{\mu\nu}}{G_0} = 8 \pi T_{\mu\nu}
\ee
Replacing $G_0 \to G_k$ requires a modification of the effective source on the right-hand side to guarantee consistency with the Bianchi identity $\nabla^\mu G_{\mu\nu} = 0$.  We can absorb this contribution into the left-hand side and write the RG-improved Einstein equations in the form 
 \be\label{eq:FieldEqEinsteinActionMod0}
 \frac{G_{\mu\nu}}{G_k} \,+\,  \frac{\Delta_k G_{\mu\nu}}{G_k}= \, 8 \pi   T_{\mu\nu} \,.
 \ee
Without the gravitational correction term $\Delta_k G_{\mu\nu}$ generated as a result the RG-improvement, the resulting field equations would be inconsistent, since the term $G_{\mu\nu}/G_k$ is not divergence-free for a generic choice of scale parameter $k$. In particular, the divergence of the extra  term has to satisfy $\nabla^\mu \Delta_k G_{\mu\nu} = \qty(\partial^\mu G_k) 8 \pi  T_{\mu\nu}$,
%$8 \pi \nabla^\mu \Delta T^k_{\mu\nu} = G_{\mu\nu} \partial^\mu G_k^{-1}$
 in order to guarantee consistency of the RG-improved Einstein quations with the Bianchi identity and the covariant conservation of the external source $T_{\mu\nu}$, without a priori constraining the form of $k$. An additional prescription to deduce the form of such an effective source 
$\Delta_k G_{\mu\nu}$ is thus necessary to define the procedure of RG-improvement at the level of the classical field equations. Later, in Section~\ref{SecSub:RGImprovementSolutions}, we will show how the  term $\Delta_k G_{\mu\nu}$ generated as a result of RG-improving the Einstein equations, can be absorbed into a redefinition of the generalized Einstein tensor for spherically symmetric spacetimes~\cite{Carballo-Rubio:2025ntd}, whose construction is reviewed in Section~\ref{SecSub:2dHorndeski}.

For completeness, let us note that the constraint imposed by the conservation of the Einstein and classical energy-momentum tensors can be accounted for by adding a  non-zero cosmological coupling $\Lambda_0$ to the Einstein equations~\eqref{eq:EinsteinFieldEq}, which itself would have to subjected to an RG-improvement $\Lambda \to \Lambda_k$. In this case, the covariant conservation of the Einstein tensor and the classical energy-momentum tensor in the RG-improved Einstein equations imposes a constraint on the allowed form of scale identification $k(x)$. Throughout this work we will however not consider a cosmological coupling.~\footnote{In general, analysing the effects of an RG-improved cosmological coupling in classical spacetimes can be further informative about conditions for singularity resolution in asymptotic safety, cf., e.g.~\cite{Adeifeoba:2018ydh}.}

Finally, it should be noted that in the case of vanishing external source in~\eqref{eq:EinsteinFieldEq}, an RG-improvement of the Newton coupling at the level of the Einstein equations~\eqref{eq:EinsteinFieldEq} is redundant, as the field equations remain those of vacuum general relativity. 

\subsection{RG-improvement of classical actions}\label{SecSub:RGActions}

In this subsection, we review the RG-improvement procedure~\eqref{eq:Gk} applied to the classical Einstein--Hilbert action,
\be\label{eq:EinsteinAction}
S_{0} = \frac{1}{16 \pi G_0}\int \dd[4]{x}\sqrt{-g}R  \,.
\ee
 The variation of the action with respect to the  inverse metric yields the left-hand side of the Einstein field equations up to a proportionality factor,
\be \label{eq:FieldEqEinstein}
\frac{G_{\mu\nu}}{G_0} = 8 \pi T_{\mu\nu}\,,
\ee
where $G_{\mu\nu} = R_{\mu\nu} - 1/2 g_{\mu\nu} R$ denotes the Einstein tensor. The energy-momentum tensor $T_{\mu\nu}$ on the right-hand side encodes the contribution from external matter sources added to the action~\eqref{eq:EinsteinAction}. Performing the prescription~\eqref{eq:Gk} in~\eqref{eq:EinsteinAction} leads to a modified action
\be\label{eq:EinsteinActionRGImproved}
S_k = \int \dd[4]{x} \sqrt{-g} \frac{R}{16 \pi G_k} \,,
\ee
resulting in modified field equations that can be written in a form analogous to~\eqref{eq:FieldEqEinsteinActionMod0} as
\be\label{eq:FieldEqEinsteinActionMod}
\frac{G_{\mu\nu}}{G_k} \,+\,  \frac{\Delta_k G_{\mu\nu}}{G_k}= \, 8 \pi   T_{\mu\nu} \,,
\ee
where, however, $\Delta_k G_{\mu\nu}$ differs from~\eqref{eq:FieldEqEinsteinActionMod0} and is given by
\ba
 \Delta_k G_{\mu\nu}=  -G_k \qty[\nabla_\mu \nabla_\nu - g_{\mu\nu}\Box]\frac{1}{8 \pi G_k} -\frac{R}{8 \pi G_k}\derivative{G_k}{k} \frac{\var k}{\var g^{\mu\nu}}\,.
\ea
Their differential order and explicit form depends on the choice of cutoff function $k$ entering the RG-improved Newton coupling~\eqref{eq:Gk}. The first term in the above equation arises for any non-constant function $k$, whereas the second one takes into account  a possible dependence of $k$ on the metric. In the case of a dependence of $k$ on metric derivatives, for example when $k$ is identified with a curvature invariant, additional terms will contribute to the equations of motion. Strictly speaking, since there is no coordinate invariant other than the spacetime dimension $d = g_{\mu\nu}g^{\mu\nu}$ that can be constructed out of the metric without derivatives, the second term in the equations above is either trivial, or else the action breaks four-dimensional diffeomorphism invariance. We choose to display such a term here, as it 
can be accounted for in the geometrodynamic framework discussed below.

Generically, a key challenge in applications of RG-improvement in asymptotic safety is the absence of a unique prescription of how to perform the scale identification.  Scale-setting procedures proposed in the context of RG-improvement at the level of the classical Einstein--Hilbert action involve conditions inferred from general covariance and the requirement of minimal scale dependence of the action~\cite{Babic:2004ev,Koch:2010nn,Domazet:2012tw}, which however do not apply in the absence of a cosmological coupling. Another proposal are scale settings based on the decoupling mechanism in effective field theories~\cite{Borissova:2022mgd}. At this stage, we shall remain agnostic to the specifics of the cutoff function. Later in Subsection~\ref{Sec:ScaleIdentification}, we will discuss allowable choices and required truncations such that the full RG-improved Einstein--Hilbert action, when reduced on spherically symmetric backgrounds, can be identified as a two-dimensional Horndeski theory.

\section{Effective geometrodynamics for RG-improved black-hole spacetimes in spherical symmetry
}\label{Sec:EffectiveGeometrodynamics}

In the following, we will focus exclusively  on an RG-improvement under the assumption of spherical symmetry, and illustrate how different implementations of RG-improvement at the level of the Einstein--Hilbert action, in the Einstein field equations, or at the level of the Schwarzschild solution, can be described by the general framework of two-dimensional Horndeski theory and generalized Einstein field equations for spherically symmetric four-dimensional spacetimes introduced in~\cite{Carballo-Rubio:2025ntd}. To that end, we  adopt the general parametrization of  four-dimensional spherically spacetimes as a warped product
\be\label{eq:gSphericalGeneral}
\dd{s}^2 = q_{ab}(x)\dd{x}^a \dd{x}^b + r(x)^2 \gamma_{ij}\dd{\theta^i}\dd{\theta^j}\,.
\ee
Here $q_{ab}(x)$ denotes a two-dimensional Lorentzian metric parametrized in terms of coordinates $x^a = \qty{x^0,x^1}$, and $r(x)$ is a scalar field. The notation $\gamma_{ij}$ is used for the Euclidean metric on the unit two-sphere, such that $\dd{\Omega}^2 =  \gamma_{ij}\dd{\theta^i}\dd{\theta^j} = \dd{\theta}^2 + \sin(\theta)^2 \dd{\phi}^2$ with coordinates $\theta^i = \qty{\theta^2,\theta^3} \equiv \qty{\theta,\phi}$.

We start with a review of the general effective geometrodynamic framework for spherically symmetric spacetimes resulting from the interpretation of two-dimensional Horndeski theories as describing the degrees of freedom of higher-dimensional warped-product spacetimes~\cite{Boyanov:2025pes,Carballo-Rubio:2025ntd}, and subsequently illustrate how this formalism can be applied in the framework of RG-improvement in spherical symmetry.

\subsection{Two-dimensional Horndeski theory and master field equations}\label{SecSub:2dHorndeski}

The two-dimensional Horndeski action is the most general diffeomorphism-invariant action for a two-dimensional metric $q_{ab}(x)$ coupled to a scalar field $r(x)$ to yield field equations containing up to second derivatives of these fields. The action can be written as~\cite{Horndeski:1974wa,Kobayashi:2011nu,Kobayashi:2019hrl}
\be\label{eq:ActionHorndeski}
S_{\text{Horndeski}}[q,r] = \int \dd[2]{x} \sqrt{-q} \mathcal{L}_{\text{Horndeski}}\,,
\ee
with Lagrangian density given by
\be\label{eq:LagrangianHorndeski}
\mathcal{L}_{\text{Horndeski}} = H_2(r,\chi) - H_3(r,\chi)\Box r + H_4(r,\chi) \mathcal{R} - 2 \partial_\chi H_4(r,\chi)\qty[\qty(\Box r)^2 - \nabla^a \nabla^b r \nabla_a \nabla_b r]\,.
\ee
Here $\nabla_a$ denotes the covariant derivative of the two-dimensional metric $q_{ab}$ with Ricci scalar $\mathcal{R}$ and moreover $\chi = \nabla_a r \nabla^a r$ is the kinetic term of the scalar field. The functions $H_i$ are general functions of $r$ and $\chi$.~\footnote{In two dimensions, the terms describing the non-minimal coupling of the scalar field to the metric, through the function $H_4$, can be absorbed by appropriately adjusting the functions $H_2$ and $H_3$, see, e.g.~\cite{Takahashi:2018yzc}. Thus, two-dimensional Horndeski theory features  in fact only two free functions. This is particularly obvious at the level of the equations of motion~\eqref{eq:geneqs1}--\eqref{eq:geneqs2}. Nevertheless, we choose to work with the overparametrized form of the action~\eqref{eq:ActionHorndeski}, as this form is particularly convenient when considering the spherical reduction of the Einstein--Hilbert action to yield a two-dimensional Horndeski action.} For future purposes, we will multiply the action in~\eqref{eq:ActionHorndeski} by $1/G_0$, where $G_0$ is the dimensionful Newton constant in four spacetime dimensions, i.e., it has canonical mass dimension $[G_0]=-2$. This is in order to simplify the later identification of the spherically reduced RG-improved Einstein--Hilbert action as a two-dimensional Horndeski action. As a result, the functions $H_i$ in the following will have canonical mass dimensions $[H_2]=0$, $[H_3]=-1$ and $[H_4]=-2$.\\

The field equations derived from~\eqref{eq:LagrangianHorndeski} by variation with respect to $q_{ab}$ and $r$ are given by
\begin{align}\label{eq:geneqs1}
	\mathscr{E}_{ab}(q,r)&\,\,\equiv \,\, \frac{1}{\sqrt{-q}} \frac{\var{\mathcal{L}}}{\var{q^{ab}}} = \bm{\beta}\nabla_a\nabla_b r-q_{ab}\left(\frac{1}{2}\bm{\alpha}+\bm{\beta}\square r \right)+\left(\partial_\chi\bm{\alpha}-\partial_r\bm{\beta} \right)\nabla_ar\nabla_br \,\, = \,\, 0
	\,,\\
	\mathscr{F}(q,r)&\,\,\equiv \,\, \frac{1}{\sqrt{-q}} \frac{\var{\mathcal{L}}}{\var{r}} = -\bm{\beta}\mathcal{R}+\partial_r\bm{\beta}\square r+\partial_r\bm{\alpha}+2\partial_\chi\bm{\beta}\left[\left(\square r\right)^2-\nabla_a\nabla_b r\nabla^a\nabla^br\right]\nonumber\\
	&-2\partial_r\left(\partial_\chi\bm{\alpha}-\partial_r\bm{\beta}\right)\chi-2\left(\partial_\chi\bm{\alpha}-\partial_r\bm{\beta} \right)\square r-2\partial_\chi\left(\partial_\chi\bm{\alpha}-\partial_r\bm{\beta}\right)\nabla_a r\nabla^a\chi \,\, = \,\, 0\,,\label{eq:geneqs2}
\end{align}
where 
\be\label{eq:AlphaBetaDef}
\bm{\alpha}(r,\chi) = H_2 + \chi \partial_r \qty(H_3 - 2 \partial_r H_4) \quad\,\,\, \text{and} \quad \,\,\, \bm{\beta} (r,\chi) = \chi \partial_\chi \qty(H_3 - 2 \partial_r H_4) - \partial_r H_4\,.
\ee
The variations $\mathscr{E}_{ab}$ and $\mathscr{F}$ satisfy the offshell generalized Bianchi identity
\be\label{eq:Bianchi}
\nabla^a \mathscr{E}_{ab} + \frac{1}{2} \mathscr{F}\nabla_b r = 0
\ee
as a result of general covariance of the Horndeski action~\eqref{eq:ActionHorndeski}.

The equations of motion for two-dimensional Horndeski theory can be used to construct the following master field equations,
\ba\label{eq:FieldEqGeneral}
\frac{\mathscr{G}_{\mu\nu}}{G_0} = 8\pi T_{\mu\nu}\,,
\ea
 on the configuration space of four-dimensional warped-product spacetimes~\eqref{eq:gSphericalGeneral}, by defining a generalization of the Einstein tensor $G_{\mu\nu}(g)$ as follows~\cite{Carballo-Rubio:2025ntd},
\ba\label{eq:GGeneralized}
\mathscr{G}_{\mu\nu}(q,r) &\equiv & \frac{1}{r^2}\mathscr{E}_{ab}\delta^a_\mu\delta^b_\nu-\frac{1}{4}r\mathscr{F}\gamma_{ij}\delta^i_\mu\delta^j_\nu\,.
\ea
This rank-two tensor possesses the defining properties of the Einstein tensor~\cite{Lovelock:1971yv}, i.e., it is symmetric, of second order in derivatives and covariantly conserved,
\be\label{eq:BianchiG}
\nabla^\mu \mathscr{G}_{\mu\nu} = 0\,,
\ee
as a result of the generalized Bianchi identity~\eqref{eq:Bianchi}. Here $\nabla_\mu$ denotes the covariant derivative of the four-dimensional metric $g_{\mu\nu}$ defined in~\eqref{eq:gSphericalGeneral}. In the following subsections we will show how the master field equations~\eqref{eq:FieldEqGeneral} can be applied in the context of RG-improvement in spherical symmetry.\\

Before we proceed, let us first make explicit how the spherically reduced Einstein--Hilbert action and the spherically symmetric Einstein field equations are included in the above description.
 
In general, restricting the Einstein--Hilbert action, or any generally covariant four-dimensional action, to the invariant sector under the action of the rotation group, and performing variations on the spherically-symmetric subspace, commutes with the corresponding reduction of the covariantly derived field equations, according to the principle of symmetric criticality~\cite{Palais:1979rca,Fels:2001rv,Frausto:2024egp}. In other words, performing variations of a covariant action $S[g]$ with respect to $g$, and subsequently imposing the spherically symmetric ansatz $g \to (q,r)$ in~\eqref{eq:gSphericalGeneral}, is equivalent to performing variations of the reduced action $S[g] \to S[q,r]$ with respect to $q$ and $r$. 

The Einstein--Hilbert action~\eqref{eq:EinsteinAction} reduced on the subspace of warped-product spacetimes~\eqref{eq:gSphericalGeneral} is given by
\begin{equation}\label{eq:iehspherical}
S_0 = \frac{\Omega_2}{16\pi G_0}\int\text{d}^2x\sqrt{-q}\,r^2\left[\mathcal{R}+\frac{2}{r^2}\left(1-\nabla_ar\nabla^ar\right)-\frac{4}{r}\square r\right]\,,
\end{equation}
where we have integrated over the angular variables to obtain a factor given by the area of the unit two-sphere $\Omega_2$. In the following we will absorb the factor of $\Omega_2/16 \pi$ into the normalization on the left-hand side. 
 
The spherically reduced Einstein--Hilbert action~\eqref{eq:iehspherical} can be identified as a two-dimensional Horndeski action~\eqref{eq:ActionHorndeski} with functions $H_i^0$ given by
\be
H^0_2(r,\chi) =  2 (1-\chi)\,, \,\,\,\quad \,\,\, \quad H^0_3(r,\chi) = 4 r\,, \,\,\,\quad \,\,\, \quad  H^0_4(r,\chi) =  r^2\,.
\ee
The expressions in~\eqref{eq:AlphaBetaDef} in this case reduce to
\ba
\bm{\alpha}_0 = 2 \qty(1-\chi) \,\,\,\quad \text{and} \,\,\,\quad 
\bm{\beta}_0= -2 r  \label{eq:AlphaBetaGR}\,,
\ea
and inserting this choice of functions into the master field equations~\eqref{eq:FieldEqEinstein} reproduces the Einstein field equations evaluated on warped-product spacetimes~\eqref{eq:gSphericalGeneral}.

\subsection{RG-improved Schwarzschild spacetimes and induced RG-improved master field equations and Horndeski actions}\label{SecSub:RGImprovementSolutions}

In the following, we will describe how static spherically symmetric  spacetimes arising from an RG-improvement of the Schwarzschild spacetime can be embedded into the formalism described above, starting with a review of static solutions to the master field equations as presented in~\cite{Carballo-Rubio:2025ntd} (see also the related discussion in~\cite{Kunstatter:2015vxa}).

To that end, we identify the scalar field in the two-dimensional Horndeski framework with one of the coordinates, and choose the remaining coordinate in the two-dimensional sector such that $q_{ab}$ is diagonal. These coordinates will be written as $\qty{x^0,x^1}=\{t,r\}$. For the purposes of describing RG-improvements in which the Newton coupling is replaced as in~\eqref{eq:Gk}, it suffices to consider configurations characterized by a single independent metric function $f_k(r)$, such that
\be
q_{ab}\dd{x}^a \dd{x}^b  = -f_k(r)\dd{t}^2 + \frac{\dd{r}}{f_k(r)}\,\,\,\quad \text{and} \quad \,\,\, r(x) = r \,.
\ee
We emphasise that this form of the line element is sufficient to describe the outcome of RG-improvement at the level of the Schwarzschild solution because this  procedure preserves the property $g_{tt} g_{rr} = -1$ of the metric. % which is tied to the condition $\partial_\chi \bm{\alpha} - \partial_r \bm{\beta} = 0$ as we shall see.
 By contrast, in an RG-improvement at the level of the spherically reduced Einstein equations or Einstein action, solutions will necessarily feature two distinct metric functions such that $g_{tt}g_{rr}\neq 1$ in general, as we will discuss explicitly.

In the above coordinates, $\chi_k = \nabla_a r \nabla^a r = f_k$ and we can hence express the four-dimensional metric~\eqref{eq:MetricSphericalSymmetry} of an RG-improved Schwarzschild spacetime as 
\be\label{eq:MetricSphericalSymmetryChi}
\dd{s^2} = -\chi_k(r)\dd{t}^2 + \frac{\dd{r}^2}{\chi_k(r)}+ r^2   \dd{\Omega}^2\,.
\ee
With these identifications, the generalized Einstein tensor~\eqref{eq:GGeneralized} evaluated on~\eqref{eq:MetricSphericalSymmetryChi} is diagonal, with $tt$ and $rr$ components given by
\ba
	\mathscr{G}^k_{tt}&=& \frac{\chi_k}{2 r^2}\left[\bm{\alpha}_k+\bm{\beta}_k\partial_r\chi_k\right]\,,\label{eq:GGentt}\\
	\mathscr{G}^k_{rr}&=&-\frac{1}{2 r^2 \chi_k^2}\qty[\bm{\beta}_k\chi_k\partial_r\chi_k+\bm{\alpha}_k \chi_k-2\chi_k^2\left(\partial_{\chi_k}\bm{\alpha}_k-\partial_r\bm{\beta}_k \right)]\label{eq:GGenrr}\,.
\ea
We do not display the angular components, as these are related to the above components by the algebraic Bianchi identity~\eqref{eq:BianchiG}. In particular, the offshell identity~\eqref{eq:Bianchi} implies that $\mathscr{F} = 0$ when $\mathscr{E}_{ab} = 0$. Therefore, solving the $tt$ and $rr$ components of the master field equations~\eqref{eq:FieldEqGeneral} in vacuum is sufficient to ensure that the angular components are satisfied. 
Setting~\eqref{eq:GGentt} and~\eqref{eq:GGenrr} to zero leaves us with the two independent equations
\ba
	\bm{\alpha}_k(r,\chi_k)+\bm{\beta}_k(r,\chi_k)\partial_r\chi_k &=&0\,,\label{eq:Eq1}\\
	\partial_{\chi_k}\bm{\alpha}(r,\chi_k)-\partial_r\bm{\beta}_k(r,\chi_k)&=&0\label{eq:Eq2}\,.
\ea

After specifying a scale identification $k^2=k^2\qty(r;M)$, such as, e.g.~in~\eqref{eq:k1pq}, the resulting RG-improved  spacetimes are described by the line element~\eqref{eq:MetricSphericalSymmetryChi} with a lapse function of the general form
\be\label{eq:ChiM}
\chi_{k}(r) = 1 - \frac{2 G_k(r;M) M}{r}\,,
\ee
where $G_k(r;M)$ denotes the effective gravitational Newton coupling. The mass parameter $M$ can more generally be seen to arise as an integration constant when solving the generalized Einstein equations~\eqref{eq:FieldEqGeneral}, as we shall see below. From the condition $G_k(r;M) \to G_0$ for $r\to \infty$, we may identify $M$ with the ADM mass of these spacetimes.

We will now show that for any metric with line element of the form~\eqref{eq:MetricSphericalSymmetryChi} and RG-improved lapse function $\chi_k(r)$ as in~\eqref{eq:ChiM}, there exist functions $\bm{\alpha}_k(r,\chi)$ and $\bm{\beta}_k(r,\chi)$ such that this spacetime is a solution to the second-order master field equations in vacuum, i.e., to Eq.~\eqref{eq:FieldEqGeneral} with $T_{\mu\nu}=0$ or, equivalently, to Eqs.~\eqref{eq:Eq1}--\eqref{eq:Eq2}. This implies in particular that these configurations arise as solutions to corresponding two-dimensional Horndeski theories. 

In view of the integrability condition represented by Eq.~\eqref{eq:Eq2}, a solution to the equations of motion can be found by means of a function $\bm{\Omega}_k(r,\chi)$~\cite{Carballo-Rubio:2025ntd,Kunstatter:2015vxa,Boyanov:2025pes}, offshell satisfying
\begin{equation}\label{eq:AlphaBetaSol}
\frac{\partial\bm{\Omega}_k(r,\chi)}{\partial r}=\bm{\alpha}_k(r,\chi)\,\,\, \quad \text{and} \,\,\,\quad \frac{\partial\bm{\Omega}_k(r,\chi)}{\partial \chi}=\bm{\beta}_k(r,\chi)\,,
\end{equation}
and evaluating to an integration constant onshell, i.e., when $\chi = \chi_k$. From their definitions in Eq.~\eqref{eq:AlphaBetaDef}, we observe that $\qty[\bm{\alpha}_0] = 0$ whereas $\qty[\bm{\beta}_0] = -1$. Therefore, the searched for function must have canonical mass dimension $\qty[\bm{\Omega}_0] =-1$.

It is reasonable to expect that $\bm{\Omega}_k$ is related to the mass. Indeed, any solution of the algebraic relation \be\label{eq:Omegimpdef}
\chi = 1 - \frac{2 G_k\left[r;\lambda \bm{\Omega}_k(r,\chi)/G_0\right] \lambda \bm{\Omega}_k(r,\chi)/G_0}{r}\,,
\ee
where $\lambda$ is a dimensionless number, results in a function $\bm{\Omega}_k$ satisfying all the above conditions. We can fix the numerical prefactor $\lambda$ so that, by convention,
\be\label{eq:OmegaOnshell}
\bm{\Omega}_k(r,\chi_k)=4G_0M\,. 
\ee
As an example, we can solve Eq.~\eqref{eq:Omegimpdef} for general relativity ($k=0$), resulting in
\begin{equation}
\bm{\Omega}_0=2r(1-\chi)\,,
\end{equation}
as well as
\be\label{eq:AlphaBetaGR2}
\bm{\alpha}_{0}=2(1-\chi) \,\,\,\quad \text{and} \quad \,\,\,\bm{\beta}_{0}=-2r\,,
\ee
which is consistent with~\eqref{eq:AlphaBetaGR}.

Another example in which Eq.~\eqref{eq:Omegimpdef} can be solved explicitly, is the case of an effective Newton coupling $G_k(r)$ with an RG scale parameter $k=k(r)$ which is independent of the mass parameter $M$, i.e., when the RG-improved lapse function is given as follows,
\ba
\chi_k(r) &=& 1 - \frac{2 G_k(r) M}{r}\,.
\ea
Such a functional dependence of the scale parameter is unphysical, as $M$ is the only physical scale of the original Schwarzschild spacetime that can and must be used for scale identification. This example thus serves only to further illustrate mathematically how a given asymptotically flat spacetime with mass $M$ can be generated as a solution to the master field equations. The solution $M = M(r,\chi_k)$ according to~\eqref{eq:OmegaOnshell} results in the offshell identification
\be \label{eq:OmegaSol}
\bm{\Omega}_k(r,\chi)=\frac{2G_0(1-\chi)r}{G_k}\,. 
\ee
We can take the derivatives of~\eqref{eq:OmegaSol} with respect to $r$ and $\chi$ to obtain $\bm{\alpha}_k$ and $\bm{\beta}_k$ according to~\eqref{eq:AlphaBetaSol},
\be\label{eq:AlphaBetaG}
\bm{\alpha}_k(r,\chi)=\frac{2 G_0(1-\chi)}{G_k^2}\qty(G_k-r\derivative{G_k}{k}\derivative{k}{r}) \quad \,\,\, \text{and} \,\,\,\quad  \bm{\beta}_{k}(r)=-\frac{2G_0r}{G_k}\,.   
\ee
It is straightforward to verify $\bm{\alpha}_{0}$ and $\bm{\beta}_{0}$ in~\eqref{eq:AlphaBetaGR2} are recovered in the limit $G_k \to G_0$.

As the final example in this subsection, we consider explicitly the spacetime obtained through an RG-improvement of the Schwarzschild spacetime by means of the scale identification~\eqref{eq:k1}, leading to the Bonanno--Reuter Hayward-type spacetime~\cite{Bonanno:2000ep}. In this case, the solution to Eq.~\eqref{eq:Omegimpdef} can be obtained by solving
\ba
\chi_k(r) &=& 1 - \frac{2 G_0 M r^2}{r^3 + G_0^2 \omega M}
\ea
for $M = M(r,\chi_k)$, which
results in the offshell identification
\be\label{eq:OmegaSol2}
\bm{\Omega}_k(r,\chi) = \frac{4\qty(1-\chi)r^3}{ \qty[2 r^2 - (1 - \chi )G_0 \omega]}\,.
\ee
We can take the derivatives of~\eqref{eq:OmegaSol2} with respect to $r$ and $\chi$ to obtain $\bm{\alpha}_k$ and $\bm{\beta}_k$ according to~\eqref{eq:AlphaBetaSol},
\ba
\bm{\alpha}_k(r,\chi) &=& -\frac{4 (\chi -1)\qty[3 G_0 \omega r^2 (\chi -1) + 2 r^4]}{\qty[G_0 \omega (\chi-1) + 2 r^2]^2}\,,\label{eq:AlphaBR} \\
 \bm{\beta}_k(r,\chi) &=& - \frac{8 r^5}{\qty[G_0 \omega (\chi-1) + 2 r^2]^2}\label{eq:BetaBR}\,.   
\ea
This explicitly establishes the RG-improved Schwarzschild spacetime as a solution to a particular two-dimensional Horndeski theory and the master field equations~\eqref{eq:FieldEqGeneral}.

 In summary, applying the above procedure allows us to derive static RG-improved spacetimes with a lapse function of the form~\eqref{eq:ChiM} as vacuum solutions to the generalized Einstein field equations~\eqref{eq:FieldEqGeneral}, written as
\ba\label{eq:GGenkActionEquationsvac}
\frac{\mathscr{G}^k_{\mu\nu}}{G_k} = \frac{G_{\mu\nu}}{G_k} + \frac{\Delta_k G_{\mu\nu}}{G_k}   = 0\,.
\ea
The functions $\bm{\alpha}_k$ and $\bm{\beta}_k$ can be obtained as described above by taking derivatives of a function $\bm{\Omega}_k$, solution to Eq.~\eqref{eq:Omegimpdef}, or concretely determined as $\Omega(r,\chi) = 4 G_0 M(r,\chi)$ following the inversion $M = M(r,\chi_k)$ for an onshell spacetime $\chi_k(r)$. The left-hand side is obtained by inserting these expressions into the variations $\mathscr{E}^k_{ab}$ and $\mathscr{F}^k$ in~\eqref{eq:geneqs1}--\eqref{eq:geneqs2}, and computing the generalized Einstein tensor $\mathscr{G}^k_{\mu\nu}$ as defined in~\eqref{eq:GGeneralized}. Therefore, not only do these equations arise as a result of RG-improving the Schwarzschild spacetime via $G_0 \to G_k$, which translates into $G_{\mu\nu} \to \mathscr{G}^k_{\mu\nu}$, but also they imply that corresponding Horndeski actions~\eqref{eq:ActionHorndeski} giving rise to the RG-improved configurations can be reconstructed explicitly, as we shall describe below.

In fact, note that the second term $\Delta_k G_{\mu\nu}$ in the equations above absorbs all new terms that are generated beyond the GR contribution $G_{\mu\nu}$, when evaluating the variations~\eqref{eq:geneqs1}--\eqref{eq:geneqs2} as a result of the $k$-dependence of the Newton coupling. Thus, the extra  term $\Delta_k G_{\mu\nu}$ generated as a result of RG-improving the Schwarzschild spacetime, can be fully absorbed into the generalized Einstein tensor $\mathscr{G}_{\mu\nu}^k$, which in turn establishes these configurations as solutions to the corresponding two-dimensional Horndeski theories. The key modifications, compared to the analogue treatment resulting in the Einstein equations of GR, are the $r$-dependence of $\bm{\alpha}_k$, and the $\chi$-dependence of $\bm{\beta}_k$, see~e.g.~Eqs.~\eqref{eq:AlphaBR}--\eqref{eq:BetaBR}, which require a non-vanishing contribution from the element $\qty[\qty(\Box r)^2 - \nabla_a \nabla_b r \nabla^a \nabla^b r]$ in the Horndeski action~\eqref{eq:LagrangianHorndeski} that generates these spacetimes as vacuum solutions to the master field equations~\eqref{eq:FieldEqGeneral}. In general, these effective terms are essential for the consistency of the modifications and the structure of vacuum solutions. We will elaborate further on the  necessity and role of such terms later.  

We finish this subsection by providing an explicit realization of two-dimensional Horndeski actions that generate static RG-improved Schwarzschild spacetimes as vacuum solutions to the master field equations~\eqref{eq:FieldEqGeneral}. Concretely, for the more general expressions for $\bm{\alpha}_k$ and $\bm{\beta}_k$ that arise in the RG-improvement of the Schwarzschild solution, we can use the definitions in Eq.~\eqref{eq:AlphaBetaDef} to obtain the associated functions $\{H^k_i(r,\chi)\}_{i=2}^4$ and, therefore, explicitly a corresponding Horndeski action. These defining relations can be solved by~\footnote{The functions $\{H_i(r,\chi)\}_{i=2}^4$ that generate the RG-improved Schwarzschild spacetimes as effective solutions to a two-dimensional Horndeski theory are not uniquely defined, as integrating by parts can change their representation relative to the Horndeski action expressed in the form~\eqref{eq:ActionHorndeski}. This is merely a reflection of the aforementioned fact that two-dimensional Horndeski theory is characterized by only two free functions, rather than three as in the parametrization of the action adopted in~\eqref{eq:ActionHorndeski}. We are here choosing a particular realization of these functions.} 
\be\label{eq:HikSpacetimes1}
H^k_2(r,\chi) = \bm{\alpha}_k(r,\chi)\,, \,\,\,\quad \,\,\, \quad H^k_3(r,\chi) =-2\bm{\beta}_k(r,\chi)\,, \,\,\,\quad \,\,\, \quad  H^k_4(r,\chi) =  -\int\text{d}r\,\bm{\beta}_k(r,\chi)\,.
\ee

This concludes our discussion of RG-improved Schwarzschild spacetimes in the effective geometrodynamic framework based on two-dimensional Horndeski theory.

\subsection{RG-improvement at the level of the spherically symmetric Einstein equations and the spherically reduced Einstein--Hilbert action}\label{SecSub:RGImprovementEquationsActions}

\begin{figure}[h!]
    \centering
    \includegraphics[scale=0.65]{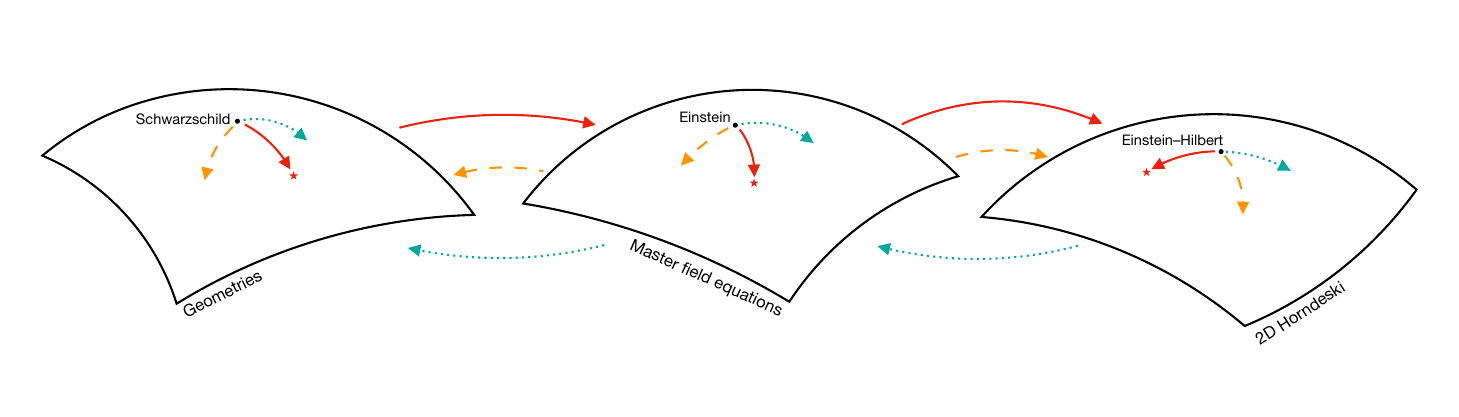}
\vspace{-0.8cm}
\caption{Schematic representation of interrelations between RG-improvement at the level of solutions, field equations and actions in spherical symmetry. From left to right, the figure depicts the sets of most general spherically symmetric geometries, the most general field equations with up to second-order derivatives of the metric, and the most general action functionals for the separate degrees of freedom defining a spherically symmetric geometry, i.e., a two-dimensional metric and a scalar field, that lead to such field equations,~i.e., the set of two-dimensional Horndeski theories. The black dots in each of these sets represent, respectively, the Schwarzschild solution, the spherically symmetric Einstein field equations, and the spherically reduced Einstein--Hilbert action. The solid red arrows represent the RG-improved Schwarzschild solution within the space of spherically symmetric geometries, as well as the induced RG-improvements within the space of master field equations and two-dimensional Horndeski actions. The red stars indicate specific examples of RG-improvements of the Schwarzschild solution, e.g., the Bonanno--Reuter black hole, for which we provide in this paper a matching set of spherically symmetric field equations and two-dimensional Horndeski action. The dashed orange arrows represent the RG-improvement implemented at the level of field equations, as well as the induced RG-improvements at the level of geometries and actions. The dotted green arrows indicate the RG-improvement implemented at the level of actions, as well as the induced RG-improvements at the level of geometries and field equations. The outcomes of these different RG-improvement procedures, indicated by the solid red, dashed orange and dotted green arrows, are in general inequivalent. The directions of the arrows connecting the sets of geometries, master field equations and two-dimensional Horndeski theories indicate the qualitative steps needed to evaluate the outcome of RG-improvement at these different levels. 
\label{fig:diagram}}
\end{figure}

In the previous subsection, we have discussed how the RG-improvement of the Schwarzschild solution results in induced RG-improvements within the space of master field equations and two-dimensional Horndeski theories. For completeness, in this subsection we discuss the possibility of implementing the RG-improvement procedure at the level of field equations and spherically reduced actions, which generally lead to different outcomes. Figure~\ref{fig:diagram} contains a summary of the following discussion.

First, we note that the RG-improvement $G_{\mu\nu} \to \mathscr{G}^k_{\mu\nu}$ associated with the RG-improvement $G_0 \to G_k$ of the Schwarzschild solution, as described in the previous subsection, is \textit{not} equivalent to the outcome of the replacement $G_0 \to G_k$ when performed at the level of the Einstein field equations. The latter induces the replacements
\ba\label{eq:AlphaBetakEinsteinEq}
\bm{\alpha}_0 \,\,\,\to \,\,\, \bm{\alpha}_k =  \frac{2G_0(1-\chi)}{G_k}\,\,\,\quad \text{and} \quad \,\,\, \bm{\beta}_0 \,\,\,\to \,\,\, \bm{\beta}_k = - \frac{2G_0 r}{G_k}\,,
\ea
for an a priori arbitrary scale identification $k = k (r,\chi)$,
which yields in general a different generalized Einstein tensor $\mathscr{G}^k_{\mu\nu}$. A key difference in the RG-improvement at the level of the classical field equations is the appearance of additional terms in the field equations, proportional to
	\ba
	\partial_\chi \bm{\alpha}_k - \partial_r {\bm{\beta}_k} &=& -\frac{2G_0 }{G_k^2}\derivative{G_k}{k}\qty[r \partial_r k + (1-\chi)\partial_\chi k] \,.
	\ea
These terms modify the structure of the corresponding vacuum  solutions to the master field equations, that generally display two independent gravitational functions and therefore go beyond the ansatz with a single function needed for RG-improvement at the level of classical solutions. Equivalently, the master field equations obtained after performing the replacements in Eq.~\eqref{eq:AlphaBetakEinsteinEq} are different from the equations that result by inserting the expressions for $\bm{\alpha}_k$ and $\bm{\beta}_k$ obtained from Eqs.~\eqref{eq:AlphaBetaSol} and~\eqref{eq:Omegimpdef}. It is important to notice that, different from an RG-improvement of the Schwarzschild solution, an RG-improvement at the level of the spherically reduced Einstein equations leads to $\partial_\chi \bm{\alpha}_k - \partial_r \bm{\beta}_k \neq 0$ in general and therefore solutions will not obey  the functional property $g_{tt}g_{rr}=-1$. The form of the metric functions can be determined once the $\bm{\alpha}_k$ and $\bm{\beta}_k$ are specified, see for instance~\cite{Carballo-Rubio:2025ntd} or the discussion in Sec.~\ref{Sec:GravCollapse} below.

Similarly,  the RG-improvement $\{H_i(r,\chi)\}_{i=2}^4 \to \{H^k_i(r,\chi)\}_{i=2}^4$ induced by the RG-improvement $G_0 \to G_k$ of the Schwarzschild solution, cf.,~e.g.~Eq.~\eqref{eq:HikSpacetimes1}, is \textit{not} equivalent to the replacement $G_0 \to G_k$ when performed at the level of the Einstein--Hilbert action. The latter  results in the expressions
\be\label{eq:Hik}
H^k_2(r,\chi) =  \frac{2G_0 (1-\chi)}{G_k} \,,\,\,\,\quad  \,\,\, \quad  H^k_3(r,\chi) = \frac{4 G_0r}{G_k}\,, \,\,\,\quad  \,\,\, \quad  H^k_4(r,\chi) =  \frac{G_0r^2}{G_k}\,,
\ee
with $G_k$ given in~\eqref{eq:Gk} and with an RG scale parameter required to depend on $k = k(r,\chi)$ in order for the construction to result in second-order field equations.~\footnote{It would be interesting to compare these expressions with constraints expected in an asymptotically safe Horndeski theory. For instance, in four dimensions it has been discussed that shift-symmetric operators are preferred over other possible terms, see~\cite{Eichhorn:2022ngh} and references therein. A detailed discussion of this issue would require analyzing two-dimensional Horndeski theory in an asymptotically safe framework.} We will discuss how this constraint on the scale-identification can be satisfied in appropriate truncations of the non-reduced RG-improved Einstein--Hilbert action later in Section~\ref{Sec:ScaleIdentification}. In this latter procedure, once the functional dependence $k=k(r,\chi) $ is taken into account, in order for the resulting RG-improved Einstein--Hilbert action to take the form of a Horndeski action~\eqref{eq:ActionHorndeski}, we need to add a term 
$ - 2 \partial_\chi H^k_4(r,\chi)\qty[\qty(\Box r)^2 - \nabla^a \nabla^b r \nabla_a \nabla_b r]$ to the RG-improved Lagrangian. The necessity for such term can be seen as a manifestation of residual effects from higher-curvature contributions generated in the process of RG-improvement.

For completeness, we show the expressions in~\eqref{eq:AlphaBetaDef} for the choice of functions $\{H_i\}_{i=2}^4$ according to~\eqref{eq:Hik},
\ba
\bm{\alpha}_k (r,\chi)& =&\frac{2G_0}{G_k}\Bigg\{1+\chi\bigg[-1+\frac{2r}{G_k}\frac{\text{d}G_k}{\text{d}k}\partial_r k-\frac{2r^2}{G_k^2}\left(\frac{\text{d}G_k}{\text{d}k}\right)^2\left(\partial_rk\right)^2+\frac{r^2}{G_k}\frac{\text{d}^2G_k}{\text{d}k^2}\left(\partial_rk\right)^2\nn\\
&{}&\quad \quad \quad \quad \quad +\frac{r^2}{G_k}\frac{\text{d}G_k}{\text{d}k}\partial^2_rk\bigg]\Bigg\}\,, \label{eq:AlphakG}\\
\bm{\beta}_k(r,\chi) &=& -\frac{2G_0r}{G_k}\Bigg\{1-\frac{r}{2G_k} \frac{\text{d}G_k}{\text{d}k}\partial_rk+\chi\Bigg[\frac{2r}{G_k^2}\left(\frac{\text{d}G_k}{\text{d}k}\right)^2\partial_rk\partial_\chi k-\frac{r}{G_k}\frac{\text{d}^2G_k}{\text{d}k^2}\partial_rk\partial_\chi k\nn\\
&{}&\quad \quad \quad \quad \quad -\frac{r}{G_k}\frac{\text{d}G_k}{\text{d}k}\partial_r\partial_\chi k \Bigg]\Bigg\}\label{eq:BetakG}\,.
\ea
These are generally different from the expressions of $\bm{\alpha}_k(r,\chi)$ and $\bm{\beta}_k(r,\chi)$ obtained in the RG-improvement of the Schwarzschild solution, which must be obtained by evaluating Eq.~\eqref{eq:AlphaBetaSol} on solutions of Eq.~\eqref{eq:Omegimpdef}. As before, since $\partial_\chi \bm{\alpha}_k - \partial_r \bm{\beta}_k \neq 0$ in general in an RG-improvement at the level of the spherically reduced Einstein action, solutions will in general possess two distinct metric functions.

This concludes our discussion of RG-improvement at the level of classical solutions, field equations and actions on the spherically symmetric subspace. 

\section{Gravitational collapse into RG-improved spacetimes} \label{Sec:GravCollapse}

In Section~\ref{SecSub:RGImprovementSolutions}, we have derived the necessary steps to construct specific instances of the master field equations and corresponding Horndeski actions that yield a given static RG-improved black-hole spacetime as an effective vacuum solution. To that end, we have provided an explicit algorithm for the reconstruction of the associated functions $\bm{\alpha}_k(r,\chi)$ and $\bm{\beta}_k(r,\chi)$, and therefrom the functions $H^k_i(r,\chi)$ defining a particular realization of the Horndeski action.

In this section, we will discuss how this construction can be extended to describe the classical dynamical gravitational collapse of matter into these static configurations. We emphasize that the following treatment is distinct from performing an RG-improvement in a fully dynamical setting. In dynamical classical spacetimes, additional scales beyond curvature and radial energy-density or pressure scales are available, which must be taken into account when performing the scale identification. See, e.g.,~\cite{Borissova:2022mgd} for an implementation making use of the decoupling scale derived from the scale-dependent effective action, and also~\cite{Delaporte:2024and} for an implementation following a different principle. Depending on the specifics of the implementation, the resulting time-dependent and $r$-dependent effective Newton coupling will in general differ from the Newton coupling derived by making a given static RG-improved spacetime time-dependent as described below. In particular, in a more complete treatment of RG-improvement for dynamical spherically symmetric spacetimes, the presence of additional matter fields relevant for scale identification can be accounted for by allowing the functions $\{H_i(r,\chi)\}_{i=2}^4$ to depend on additional variables. Related, in the presence of a scalar matter field $\phi$, one may consider the extension to functions $\{H_i(r,\chi,\phi,\nabla_\mu\phi\nabla^\nu\phi)\}_{i=2}^4$, resulting in a biscalar Horndeski theory~\cite{Ohashi:2015fma,Nejati:2024tuo}.~\footnote{Constructing the most general Lagrangians extending Horndeski's construction to an arbitrary number of scalar fields is an open problem~\cite{Katayama:2025hnd}.} Describing the impact of additional matter scales in RG-improved gravitational collapse processes, as well as the resulting spacetimes, is beyond the scope of this paper. Here, we will instead focus on using classical matter sources on the right-hand side of the Einstein and master field equations, that can achieve the classical gravitational collapse into an RG-improved Schwarzschild spacetime as the endpoint of dynamical evolution, without introducing any additional physical scales in the definition of the RG scale parameter.~\footnote{This analysis can thus further be informative about the role that matter scales, in addition to the gravitational scales, may play, in what concerns the outcome of a quantum-gravitational collapse.}

To begin, we consider the RG-improved master field  equations~\eqref{eq:GGenkActionEquationsvac}, with the RG-improved generalized Einstein tensor on the left-hand side, in combination with a classical source on the right-hand side, i.e.,
\ba
\frac{\mathscr{G}_{\mu\nu}^k}{G_k} &=& 8 \pi T_{\mu\nu}\,.
\ea
To describe the dynamical RG-improved Schwarzschild spacetimes $f_k$, we consider a new set of coordinates $\qty{x^0,x^1}=\qty{v,r}$ for the two-dimensional metric $q_{ab}$, such that
\be \label{eq:met1}
q_{ab}\dd{x}^a \dd{x}^b  = -F_k(v,r)\dd{v}^2 + 2 H_k(v,r) \dd{v}\dd{r}\,\,\,\quad \text{and} \quad \,\,\, r(x) = r \,.
\ee
As a result, changing the notation for the scalar-field kinetic term to $X_k = \nabla_a r \nabla^a r = F_k/H_k^2$, we will express the four-dimensional metric~\eqref{eq:MetricSphericalSymmetry} of a dynamical spacetime, to be evolved into a static RG-improved spacetime, as 
\be\label{eq:MetricSphericalSymmetryChiDynamical}
\dd{s^2} = -X_k(v,r) H_k(v,r)^2\dd{v}^2 + 2 H_k(v,r)\dd{v}\dd{r}+ r^2   \dd{\Omega}^2\,.
\ee
Note that $\qty{v,r,\theta^2,\theta^3}$ acquire the interpretation of advanced Eddington-Finkelstein coordinates.

We proceed for concreteness by considering  a time-dependent Vaidya source~\cite{Vaidya:1951zza,Vaidya:1966zza} corresponding to a pressureless perfect fluid, in the form
\ba\label{eq:FieldEqGeneralDynamical}
T_{\mu\nu} = \frac{\dot{m}(v)}{4\pi r^2} \partial_\mu v \partial_\nu v\,,
\ea
where a dot denotes the derivative with respect to $v$. The solutions of the master field equations for such a source, and with $H_k(v,r)=1$, were presented in~\cite{Boyanov:2025pes}. While solutions with $H_k(v,r)=1$ would be sufficient to describe gravitational collapse processes into the RG-improved spacetimes described in Section~\ref{SecSub:RGImprovementSolutions}, considering more general functions $H_k(v,r)$ is necessary in order to include the spacetimes resulting from the treatment in Section~\ref{SecSub:RGImprovementEquationsActions}. %\change{[The static versions of these spacetimes can be obtained straightforwardly by imposing $\dot{m}=0$ below.]}

Note that $\mathscr{G}^k_{\mu\nu}$ on the left-hand side now denotes the generalized Einstein tensor~\eqref{eq:GGeneralized} computed for the dynamical spacetime~\eqref{eq:MetricSphericalSymmetryChiDynamical}. It has three independent components given by
\ba 
\mathscr{G}_{vv}^k &=& \frac{H_k}{2 r^2 }\qty[X_k H_k \qty(\bm{\alpha}_k + \bm{\beta}_k  \partial_rX_k) + \bm{\beta}_k \partial_v X_k]\label{eq:Eq1Dynamical}\,,\\
\mathscr{G}_{vr}^k &=& -\frac{H_k}{2 r^2}\qty[\bm{\alpha}_k + \bm{\beta}_k \partial_r X_k ] \label{eq:Eq2Dynamical}\,,\\
\mathscr{G}_{rr}^k &=& \frac{1}{r^2 H_k}\qty[H_k \qty(\partial_{X_k}\bm{\alpha}_k - \partial_r \bm{\beta}_k) - \bm{\beta}_k \partial_r H_k]\label{eq:Eq3Dynamical} \,,
\ea
where $\bm{\alpha}_k = \bm{\alpha}_k(r,X_k)$ and $\bm{\beta}_k=\bm{\beta}_k(r,X_k)$. Note that for $H_k =1$ and $X_k(v,r) =  \chi_k(r)$, requiring the above expressions to vanish, the first and second equation are equivalent and we recover the two  vacuum equations of motion~\eqref{eq:Eq1}--\eqref{eq:Eq2} in the static setup. More generally, these expressions allow us to illustrate more clearly some of the previously made statements. Let us assume that no source is present for now. If the condition $\partial_\chi \bm{\alpha}_k - \partial_r \bm{\beta}_k = 0$ is satisfied, it follows that $\partial_r H =0$ such that $H$ can depend only on $v$, but this $v$-dependence can be absorbed in a redefinition of $v$ and therefore solutions can be gauge-fixed such that $g_{vr}=1$ or equivalently $g_{tt}g_{rr}=-1$ in Schwarzschild coordinates. On the other hand, if $\partial_\chi \bm{\alpha}_k - \partial_r \bm{\beta}_k \neq 0$ then this expression sources $\partial_r H$ such that solutions in this case will generically feature two distinct metric functions. This is the case in an RG-improvement at the level of the spherically reduced Einstein equations or Einstein action.

Now taking into account a Vaidya source, observe that the only non-zero component of the Vaidya energy-momentum tensor~\eqref{eq:FieldEqGeneralDynamical} is the $vv$-component. In particular, the expressions in~\eqref{eq:Eq2Dynamical} and~\eqref{eq:Eq3Dynamical} are set to zero by the generalized Einstein equations. From the latter we can obtain the function $H_k$, whereas using the former in Eq.~\eqref{eq:Eq1Dynamical} determines the time derivative of $X_k$. Altogether, the following three equations characterize the general solution,
\ba
H_k &=& c(v) e^{\int \dd{r} \frac{\partial_{X_k}\bm{\alpha}_k - \partial_r \bm{\beta}_k}{\bm{\beta}_k}}\,,\label{eq:EqH}\\
\bm{\alpha}_k+\bm{\beta}_k\partial_r X_k &=& 0\label{eq:EqXR}\,,\\
 \partial_v X_k &=&  \frac{4 G_0\dot{m}}{H_k \bm{\beta}_k} \label{eq:EqXDot}\,, 
\ea
where $c(v)$ can be absorbed in a redefinition of $v$. In the limit $\dot{m} \to 0$ we see that $X_k$ becomes time-independent, as expected. A direct computation shows that the condition $\partial_r\partial_v X_k=\partial_v\partial_rX_k$ follows from these equations, which guarantees their integrability. Notice moreover, that when $\dot{m}=0$, these expressions state the generic solutions in an RG-improvement at any of the three levels --- Schwarzschild solution, spherically reduced Einstein equations, spherically reduced Einstein action --- whereby $H=1$ can be achieved only in the first case when $\partial_\chi \bm{\alpha}_k - \partial_r \bm{\beta}_k = 0$.

Specific dynamical solutions $X_k(v,r)$ and $H_k(v,r)$  can be found once a given functional form of $\bm{\alpha}_k$ and $\bm{\beta}_k$ is specified. These two  functions encode the modifications arising from the RG-improvement of the gravitational part of the action, independently of any in addition introduced matter degrees of freedom. On these grounds, it is reasonable to impose  $\bm{\alpha}_k$ and $\bm{\beta}_k$ to be given by the analogue quantities for static RG-improved spacetimes.

Let us first as a concrete example consider the dynamical evolution into the Schwarzschild spacetime, for which $\bm{\alpha}_0$ and $\bm{\beta}_0$ must be chosen according to~\eqref{eq:AlphaBetaGR2} with $\chi_0 \to X_0$. Since these functions satisfy $\partial_{X_k}\bm{\alpha}_0 - \partial_r \bm{\beta}_0 = 0$, from~\eqref{eq:EqH} we infer $H_k = 1$. Therewith, Eqs.~\eqref{eq:EqXR} and~\eqref{eq:EqXDot} are solved by
\ba
X_0(v,r) = 1 - \frac{2 G_0 m(v)}{r}\,,
\ea
which represents the classical Vaidya spacetime~\cite{Vaidya:1951zza,Vaidya:1966zza}. For $m(v) = M$ we recover the static Schwarzschild spacetime.

\begin{figure}[t]
	\centering
	\hspace{1.8cm}
	\includegraphics[width=0.75\textwidth]{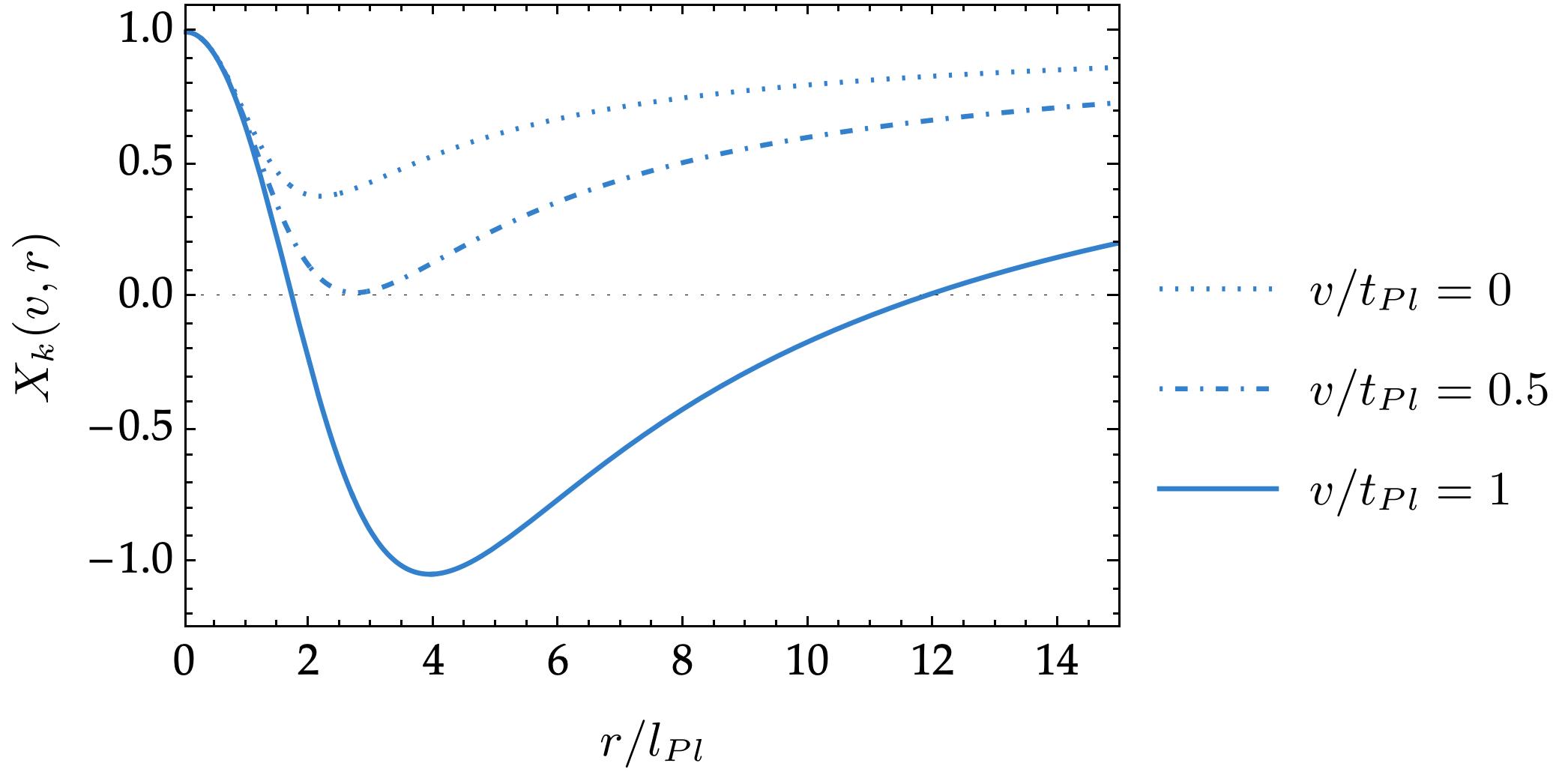}
	\caption{\label{Fig:XkBR} Gravitational collapse into a regular black-hole spacetime with time-time component of the metric $X_k(v,r)$ given by~\eqref{eq:XkSol} with a linear mass function $m(v) \propto v$, at different constant-$v$ slices. The free parameters are set to $M/m_{Pl}=1$, $G_0/m_{Pl}^{-2}=1$ and $\omega =5$.}
\end{figure} 

Next, we consider as an example the dynamical evolution into the Bonanno--Reuter Hayward-type spacetime~\cite{Bonanno:2000ep}, for which $\bm{\alpha}_k$ and $\bm{\beta}_k$ must be chosen according to~\eqref{eq:AlphaBR}--\eqref{eq:BetaBR} with $\chi_k \to X_k$. In this case, these functions still satisfy $\partial_{X_k}\bm{\alpha}_k - \partial_r \bm{\beta}_k = 0$, such that from~\eqref{eq:EqH} we infer $H_k = 1$.  Moreover, Eq.~\eqref{eq:EqXR}  has a general solution in the form
\ba
X_k(v,r) = 1 - \frac{1}{r} \frac{1}{\tau(v) + \frac{G_0 \omega}{2 r^3}}\,.
\ea
Inserting this result into Eq.~\eqref{eq:EqXDot}, we obtain the differential equation
\ba
2 G_0 \tau^2 \dot{m} + \dot{\tau} = 0
\ea
which determines the function $\tau(v)$. The solution is given by
\ba
\tau(v) = \frac{1}{\tau_0 + 2 G_0 m(v)}\,,
\ea
where $\tau_0$ is an integration constant. In summary, we obtain the dynamical solution 
\ba\label{eq:XkSol}
X_k(v,r) = 1 - \frac{1}{r}\qty(\frac{1}{2 G_0 \qty(M + m(v))}+ \frac{G_0\omega}{2 r^3})^{-1}\,,
\ea
where we have the fixed integration constant such that the static RG-improved spacetime is recovered in the limit $m(v)\to 0$. Figure~\ref{Fig:XkBR} shows $X_k(r,v)$ for a linear choice of mass function $m(v) \propto v$ at different constant-$v$ slices. 

\section{Connecting 2D Horndeski theory and 4D generally covariant actions}\label{Sec:ScaleIdentification}

In Section~\ref{SecSub:RGImprovementSolutions}, we have established that static RG-improved spacetimes can be understood as vacuum solutions to the master field equations~\eqref{eq:FieldEqGeneral} for specific choices of functions ${H_i}$ appearing in the two-dimensional Horndeski action. In addition, in Section~\ref{SecSub:RGImprovementEquationsActions} we have seen that the RG-improvement of the spherically reduced Einstein--Hilbert action amounts to the multiplication of the corresponding functions $H_i$ by $G_0/G_k$, where the scale parameter in the RG-improved Newton coupling is assumed  to be a function $k = k(r,\chi)$. Both outcomes are in general not manifestly connected to a four-dimensional generally covariant action. Here, we will discuss how the corresponding two-dimensional Horndeski theories can arise in appropriate truncations of four-dimensional generally covariant actions. 

For concreteness, we start by asking the question of what types of Horndeski theories can arise by truncating a general four-dimensional local action
 \ba\label{eq:S}
 S[g] &=& \int \dd[4]{x}\sqrt{-g}\mathcal{L}\qty(g^{ab},R^{(4)}_{abcd})\,,
 \ea
 built from scalar invariants of the Riemann tensor without covariant derivatives, to the sector yielding second-order field equations when reduced on warped-product backgrounds~\eqref{eq:gSphericalGeneral}. We consider here specifically $f(\text{Riemann})$ theories as a large enough class of theories that allows us to make non-trivial statements. This is also the relevant class of theories to make connections with the discussion of polynomial quasi-topological gravities in higher dimensions~\cite{Bueno:2025qjk}, whose reduced actions are two-dimensional Horndeski actions of the type arising from our truncation procedure described below. In a companion paper~\cite{Borissova:2026wmn} we will establish this connection explicitly and show that all of the two-dimensional Horndeski actions which can arise in truncations described below, can in fact arise in consistent reductions of four-dimensional non-polynomial gravities such that the spacetimes which can be generated from such Horndeski theories can be obtained as four-dimensional gravitational vacuum solutions. The extension to gravitational actions involving covariant derivatives, as expected in a generic effective field theory of gravity~\cite{Ruhdorfer:2019qmk}, and analyses of the thereby implied extension of the solution space will be addressed in future works.
 
To answer the above raised question, we observe that the Riemann tensor evaluated for this class of spacetimes takes the form
 \begin{equation}
 	R^{(4)}_{abcd}=\mathcal{R}_{abcd}\,,\quad\,\,\, R^{(4)}_{aibj}=-r\gamma_{ij}\nabla_a\nabla_br\,,\quad\,\,\, R^{(4)}_{ijkl}=r^2\left(1-\chi\right)\left(\gamma_{ik}\gamma_{jl}-\gamma_{il}\gamma_{jk}\right)\,, 
 \end{equation}
 where $\mathcal{R}_{abcd}$ denotes the Riemann tensor for the two-dimensional metric $q_{ab}$. Thus, all scalar polynomial Riemann invariants for this class of spacetimes will be polynomial functions of the four combinations
 \begin{equation}
 	\left\{\mathcal{R}\,,\,\frac{1}{r^2}\nabla_a \nabla_b r \nabla^a \nabla^b r\,,\,\frac{1}{r}\square r\,,\,\frac{1-\chi}{r^2}\right\} \,.
 \end{equation}
Taking into account that $\sqrt{-g} \to r^2$ up to an irrelevant angular constribution, we see that the truncation of the action~\eqref{eq:S} reduced on~\eqref{eq:gSphericalGeneral}, to a Horndeski action~\eqref{eq:ActionHorndeski}, will result in a restricted class of Horndeski actions, for which
\ba\label{eq:hi}
h_2(r,\chi) \,\, =\,\, \frac{H_2(r,\chi)}{r^2}\,,\quad \quad h_3(r,\chi) \,\, =\,\, \frac{H_3(r,\chi)}{r} \,, \quad \quad h_4(r,\chi) \,\, =\,\, \frac{H_4(r,\chi)}{r^2}
\ea
are functions of
\ba
\psi &\equiv & \frac{1-\chi}{r^2}\,,
\ea
i.e., the $h_i$ can depend on $r$ and $\chi$ only through the combination $\psi(r,\chi)$. This constraint is restrictive from the viewpoint of generic two-dimensional Horndeski actions. One should therefore generically not expect that generic RG-improved Schwarzschild spacetimes can arise from a generally covariant four-dimensional action of the form~\eqref{eq:S}, truncated in the described way. However, any RG-improved Schwarzschild spacetime for which the functions $\bm{\alpha}_k$ and $\bm{\beta}_k$ can be written as
\ba \label{eq:AlphaBetaConstraint}
\bm{\alpha}_k &=& r^2 \bm{a}_k(\psi) \,\,\,\quad \text{and} \quad \,\,\, \bm{\beta}_k \,\,=\,\, r \bm{b}_k(\psi)\,,
\ea
{\it can} be obtained from such a truncation. The corresponding generating function $\bm{\Omega}$ must be of the form
\begin{equation}\label{eq:OmegaConstraint}
\bm{\Omega}_k=r^3\bm{\omega}_k(\psi).    
\end{equation}

On the other hand, from the definition in~\eqref{eq:AlphaBetaDef}, considering the change of variables
\ba
(r,\chi)\quad \to \quad  (r,\psi)\,, \quad \quad 
(\partial_r,\partial_\chi)  \quad \to\quad \qty(\partial_r-\frac{2\psi}{r}\partial_\psi, -\frac{1}{r^2}\partial_\psi)\,,
\ea
we infer that the constraint $h^k_i = h^k_i(\psi)$ and the definitions in Eq.~\eqref{eq:AlphaBetaDef} imply that the functions $\bm{\alpha}_k$ and $\bm{\beta}_k$ take the form
\ba
\bm{\alpha}_k(r,\psi) &=&  r^2 h^k_2(\psi) - \qty(1-r^2 \psi) \qty[- h^k_3(\psi) + 2 \psi {h^k_3}'(\psi) + 4 h^k_4(\psi) - 4 \psi {h^k_4}'(\psi) + 8 \psi^2 {h^k_4}''(\psi)]\,,\,\,\,\\
\bm{\beta}_k(r,\psi) &=&  
- \frac{1}{r}\qty(1- r^2 \psi) \partial_\psi \qty[h^k_3(\psi) - 4 h^k_4(\psi) + 4 \psi {h^k_4}'(\psi)] - 2 r h^k_4(\psi) + 2 r \psi {h^k_4}'(\psi)\,.
\ea
Comparing with~\eqref{eq:AlphaBetaConstraint}, we can find a representation for the functions $h_i^k(\psi)$ as follows,
\be\label{eq:hikSpacetimes}
h^k_2(\psi) = \bm{a}_k(\psi)\,, \,\,\,\quad \,\,\, h^k_3(\psi) \,\,= \,\,4 h^k_4(\psi)-4\psi {h^k_4}'(\psi)=-2\bm{b}_k(\psi)\,, %\,, \,\,\,\quad \,\,\, \quad  h^k_4(\psi) = \frac{1}{2}\bm{b}_k(\psi) \,,
\ee
i.e., such that 
\begin{equation}
 h^k_4(\psi) =\frac{1}{2}\psi\int\text{d}\psi\,\frac{\bm{b}_k(\psi)}{\psi^2}\,,    
\end{equation}
which is a solution of the differential equation in~\eqref{eq:hikSpacetimes}. The resulting representation of the functions $H^k_i$ thus takes the form
\be\label{eq:HikSpacetimes2}
H^k_2(r,\chi) = r^2\bm{a}_k(\psi)\,, \,\,\,\quad \,\,\, \quad H^k_3(r,\chi) =-2r\bm{b}_k(\psi)\,, \,\,\,\quad \,\,\, \quad  H^k_4(r,\chi) =\frac{1}{2} r^2\psi\int\text{d}\psi\,\frac{\bm{b}_k(\psi)}{\psi^2}\,.
\ee

This result is equivalent to the representation given previously in~\eqref{eq:HikSpacetimes1} for $\bm{\beta}_k(r,\chi)=r\bm{b}_k(\psi)$. Indeed, the difference between $H_4^k(r,\chi)$ in both representations satisfies
\begin{equation}\label{eq:repseqchi}
	\partial_r\left[-\int\text{d}r\,r\bm{b}_k(\psi) -\frac{1}{2} r^2\psi\int\text{d}\psi\,\frac{\bm{b}_k(\psi)}{\psi^2}\right]=-r\bm{b}_k(\psi)-\frac{1}{2}r^2\psi\frac{\bm{b}_k(\psi)}{\psi^2}\left(-\frac{2\psi}{r}\right)=0\,,   
\end{equation}
where $\psi=\psi(r,\chi)$ and $\partial_r$ is the partial derivative with respect to $r$ in the variables $(r,\chi)$. Thus, from~Eq.~\eqref{eq:AlphaBetaDef} we infer that the difference between $H^4_k(r,\chi)$ in both representations does not contribute to the equations of motion. 
Hereby we conclude that the associated RG-improved Schwarzschild spacetimes can arise from the truncation of the local action~\eqref{eq:S} to the sector that leads to second-order equations of motion upon evaluation on~\eqref{eq:gSphericalGeneral}. Such a truncation can only be defined on the space of warped-product spacetimes~\eqref{eq:gSphericalGeneral}.

Specific examples include the Bonanno--Reuter spacetime defined through~\eqref{eq:AlphaBR}--\eqref{eq:BetaBR}, i.e., for which
\ba \label{eq:AlphaBetaConstraintConcrete}
\bm{a}_k(\psi) &=& - \frac{4 \psi \qty( 3 G_0 \omega \psi -2 )}{\qty(G_0 \omega \psi - 2)^2}  \,\,\,\quad \text{and} \quad \,\,\, \bm{b}_k(\psi) \,\,=\,\, -\frac{8}{\qty(G_0 \omega \psi - 2)^2}\,,
\ea
and
\ba
\bm{\omega}_k(\psi) &=& \frac{4\psi}{2 - G_0 \omega \psi}\,.
\ea
The Dymnikova spacetime~\cite{Dymnikova:1992ux}, which has arisen in previous discussions of RG-improvement~\cite{Platania:2019kyx}, see also~\cite{Borissova:2022mgd} for variants thereof, also satisfies these requirements, even though the solution to Eq.~\eqref{eq:Omegimpdef} cannot be written in terms of elementary functions. Other regular black holes proposed in phenomenological and numerical explorations of regular gravitational collapse~\cite{Ziprick:2010vb} do not satisfy the requirements in Eqs.~\eqref{eq:AlphaBetaConstraint} and~\eqref{eq:OmegaConstraint}, and therefore cannot be connected to an action principle of the form in Eq.~\eqref{eq:S}.

In the final part of this section, we proceed to consider a special type of truncation required in order for a generic RG-improved Einstein--Hilbert action to yield a Horndeski action. To that end, we remind that the RG-improved Einstein--Hilbert action is obtained by the replacement $G_0 \to G_k$ with $G_k$ depending on the RG scale parameter $k$ as in~\eqref{eq:Gk}. In particular, it will be a non-polynomial function of curvature invariants entering the RG scale parameter $k$. Generically, $k$ can be a function of curvature invariants of the metric, as well as of external matter fields such that four-dimensional diffeomorphism invariance is preserved. Here we exclude a dependence on external scales, as we are only interested in actions that may produce  a two-dimensional Horndeski action of the type~\eqref{eq:ActionHorndeski} for an appropriately constrained choice of the scale parameter $k=k(r,\chi)$. 

This constraint can be satisfied by a systematic truncation procedure of curvature invariants defined as follows.
Let us initially assume that the RG scale parameter $k^2$ is identified with a four-dimensional scalar polynomial invariant $I^{(4,2n)}$ built from $n$ powers of the Riemann tensor and thus involving $2n$ derivatives. On dimensional grounds, we may identify $k^2$ with the $n$th root of an such invariant, or alternatively multiply $I^{(4,2n)}$ with $2n-2$ powers of a mass scale to ensure that $k^2$ is mapped onto a quantity with canonical mass dimension two, i.e., $\qty[k^2] = +2$. The constraint demanding a functional dependence $k = k(r,\chi)$ can be satisfied by noticing that in any Riemann invariant $I^{(4,2n)}$ evaluated for the spacetimes~\eqref{eq:gSphericalGeneral}, the only contribution that does not introduce any derivatives of $q_{ab}$, or second-order derivatives of the scalar field $r$, arises from the purely angular components of the four-dimensional Riemann tensor. In other words, we are forced to truncate the dependence of $G_k$ in the RG-improved Einstein action on curvature invariants as follows,
\ba\label{eq:Gktruncation}
G_k = \frac{G_0}{1 + G_0 \omega k^2\qty(I^{(4,2n)}\qty[R^{(4)}_{abcd},R^{(4)}_{aibj},R^{(4)}_{ijkl}])} \,\,\, \quad \to \quad \,\,\,  G_k = \frac{G_0}{1 + G_0 \omega k^2\qty(I^{(4,2n)}\qty[0,0,R^{(4)}_{ijkl}])}\,.\,\,\,
\ea
For any curvature invariant $I^{(4,2n)}$ built from $n$ powers of the Riemann tensor, the truncation defined by~\eqref{eq:Gktruncation} amounts to
\ba\label{eq:RiemN}
I^{(4,2n)} \,\,\,\to \,\,\, \psi^n \,\,= \,\, \frac{(1-\chi)^n}{r^{2n}}\,,
\ea
up to a proportionality constant. This truncation of four-dimensional curvature invariants is manifestly invariant under two-dimensional diffeomorphisms and guarantees that the resulting action describes a two-dimensional Horndeski theory in which $G_k = G_k(\psi)$, regardless of the specific combination of curvature invariants used for scale identification in the original RG-improved action. In particular, it allows us to retain non-trivial partial contributions from an infinite series of curvature invariants. Note, concretely, that the resulting functions $H_i$ given in~\eqref{eq:Hik} take the form
\ba
H_2^k(r,\chi) \,\,= \,\, \frac{2 G_0 \qty(1-\chi)}{G_k(\psi)}\,,\quad \quad H_3^k(r,\chi) \,\, = \,\, \frac{4 G_0 r}{G_k(\psi)} \,, \quad \quad H_4^k(r,\chi) \,\, =\,\, \frac{G_0 r^2}{G_k(\psi)}\,,
\ea
and thus automatically acquire the functional dependence ensuring $h^k_i = h^k_i(\psi)$.

\section{Discussion}\label{Sec:Discussion}

RG-improvement has been developed as a qualitative pathway for analysing the implications of a UV point for quantum gravity, on the regularity, global structure and final state of black-hole evaporation~\cite{Eichhorn:2022bgu,Platania:2023srt}. The weakening of the dimensionful Newton coupling implied by  quantum scale symmetry provides a mechanism for the resolution of classical singularities and may result in regular black holes or horizonless black-hole mimickers as candidate alternatives towards a non-singular paradigm of black-hole physics~\cite{Carballo-Rubio:2025fnc}. Yet, the heuristic application of RG-improvement is ambiguous in the freedom of implementation at the level of the action, in field equations or solutions characterizing a classical theory, as well as in the procedure of scale identification. Here, we have established relations between these different notions and outcomes of RG-improvement in spherical symmetry, using the framework of the master field equations for spherically symmetric gravitational fields, recently constructed from two-dimensional Horndeski theory~\cite{Carballo-Rubio:2025ntd}. We have shown how effective spacetimes arising from an RG-improvement of the Schwarzschild spacetime, such as the original Bonanno--Reuter regular black-hole spacetime~\cite{Bonanno:2000ep}, can be viewed as vacuum solutions to these master field equations. These spacetimes can therefore be understood as solutions to a generally covariant two-dimensional Horndeski theory, whereby the corresponding action can be explicitly reconstructed. Moreover, we have demonstrated how the effects of higher-order curvature contributions, arising in an RG-improvement of the action, can be  accounted for while preserving the second-order nature of the resulting field equations. 

We view these analyses as a systematic and operational approach to RG-improvement, in which previous works can be included and extended. Our work focuses on providing a mathematical framework for RG-improvement, without delving into the validity of the latter as a procedure of qualitatively capturing quantum-gravity corrections to classical spacetimes in a controlled manner. In fact, our work establishes manifestly the discrepancies of the outcomes of RG-improvement implemented at different levels in a classical theory. Moreover our work focuses on the highly symmetry-reduced framework of spherically symmetric black-hole spacetimes, whereas astrophysical black holes are expected in general to have non-zero spin. In this case quantum-gravity effects can be qualitatively addressed in an RG-improvement of the Kerr metric~\cite{Held:2019xde,Eichhorn:2021etc,Delaporte:2022acp}. Our discussion is not directly applicable to spacetimes with non-zero spin and extending our arguments to include these situations is non-trivial.
%The extension of our discussions to spacetimes with non-zero spin is non-trivial as they exploit crucially the issue of consistent truncation and principle of symmetric criticality which applies for compact symmetry groups. It is at this point not clear if or how our discussion could be extended to non-static stationary spacetimes. 
Possible related directions to explore are whether the Kerr offshell families discussed in~\cite{BenAchour:2025uzp} can be generated from an action principle, or the construction of slowly rotating black holes in quasi-topological gravity~\cite{Fierro:2020wps}.

With these remarks, we have discussed how to construct effective field equations and actions for a given choice of RG scale parameter such that the corresponding static RG-improved black-hole spacetimes arise as vacuum solutions. This also allowed us to discuss the classical dynamical gravitational collapse of matter into these static configurations. Among possible future directions of the formalism developed here, it would be interesting to consider a fully dynamical treatment of RG-improvement in gravitational collapse processes, in which the additional energy scales provided by the infalling matter, as well as additional backreaction effects stemming from the quantum-gravitational self-energy generated as a result of the RG-improvement, are taken into account. This can be done systematically by identifying the RG scale parameter with the decoupling scale, in general a combination of curvature and matter scales, and extending the procedure to obtain self-consistent spacetimes as fixed points of a series of iterated RG-improvements as in~\cite{Platania:2019kyx,Borissova:2022mgd}. From the perspective of two-dimensional Horndeski theories, this is equivalent to allowing non-minimal couplings of matter fields to the gravitational degrees of freedom, with a subsequent increase in complexity of the analysis.

Our approach may in addition reveal relations to quasi-topological gravities in higher dimensions~\cite{Oliva:2010eb,Myers:2010ru,Dehghani:2011vu,Cisterna:2017umf,Bueno:2019ltp,Bueno:2019ycr, Bueno:2022res,Bueno:2024dgm,Bueno:2024zsx,Bueno:2025qjk}. Recent analyses of conditions for covariant densities to be of quasi-topological type, show that the corresponding gravitational actions reduced on spherically symmetric backgrounds describe particular subclasses of two-dimensional Horndeski actions parametrized entirely by a single function $h$ of the quantity $\psi$ defined in~\eqref{eq:RiemN}~\cite{Bueno:2025qjk}. The resulting static spherically symmetric solutions are necessarily single-function spacetimes, whereby the metric function $f$ is determined through an algebraic equation relating the characteristic function $h(\psi)$ to the ADM mass $M$ of the spacetime in the form $h(\psi) \propto M/r^{d-1}$~\cite{Bueno:2025qjk}. In view of our reconstruction algorithm provided in Subsection~\ref{SecSub:RGImprovementSolutions}, which involves the inversion of the algebraic relation ${\bm \Omega}(r,f) = 4 G_0 M$, together with the constrained form of the function $\bm{\Omega}(r,\psi) \sim r^3 \bm{\omega}(\psi)$ discussed in Section~\ref{Sec:ScaleIdentification}, it is reasonable to expect that the underlying spaces of two-dimensional Horndeski actions are related. In particular, by resorting to the truncation procedure of generally covariant actions to the sector yielding second-order equations of motion on spherically symmetric spacetimes described in Section~\ref{Sec:ScaleIdentification}, all the four-dimensional analogues of regular black-hole spacetimes derived from higher-dimensional quasi-topological gravities~\cite{Bueno:2024dgm,Frolov:2024hhe} can be generated. However, at the same time, our results here show that the more general subclass of Horndeski actions whose functions $h_i(\psi)$ cannot be parametrized in terms of a single function $h(\psi)$, and which can be reached by spherically reducing non-polynomial gravities, will in general give rise to static spherically solutions featuring two distinct metric functions. We will elaborate on these statements elsewhere, and finish by concluding that the effective geometrodynamic framework based on two-dimensional Horndeski theory may yield further insights into the structure and order of curvatures required for effective Lagrangian densities to regularize classical singular black holes.~\footnote{It should be emphasized that this is an onshell statement, regardless of the fact that classical singular black holes may be effectively suppressed offshell in the quantum-gravitational path integral~\cite{Borissova:2020knn,Borissova:2023kzq}.}

\section*{Acknowledgements}

The authors are grateful to Astrid Eichhorn for detailed feedback on a previous version of this manuscript. J.B.~is supported by STFC Consolidated Grant ST/X000575/1 and Simons Investigator Award~690508. R.C-R.~acknowledges financial support provided by the Spanish Government through the Ram\'on y Cajal program (contract RYC2023-045894-I) and the Grant No.~PID2023-149018NB-C43 funded~by MCIN/AEI/10.13039/501100011033, and by the Junta de Andaluc\'{\i}a 
through the project FQM219 and from the Severo Ochoa grant 
CEX2021-001131-S funded by MCIN/AEI/ 10.13039/501100011033.
 
\bibliographystyle{jhep}

\bibliography{refs}

\end{document}